\documentclass[
 reprint,
 amsmath,amssymb,
 aps,
]{revtex4-2}

\usepackage{graphicx}
\usepackage{bm}

\usepackage[colorlinks=true, linkcolor=blue,anchorcolor=blue, citecolor=blue,urlcolor=blue]{hyperref}

\usepackage[normalem]{ulem}
\begin{document}
\bibliographystyle{apsrev4-1}
\preprint{APS/123-QED}

\title{Tuning the magnetic anisotropy and topological phase with electronic correlation in single-layer H-FeBr$_2$}

\author{Weiyi Pan}%
\email{pwy20@mails.tsinghua.edu.cn}
\affiliation{%
State Key Laboratory of Low Dimensional Quantum Physics and Department of physics, Tsinghua University, Beijing 100084, China
}%

\date{\today}

\begin{abstract}
Electronic correlation can strongly influence the electronic properties of two-dimensional (2D) materials with open d- or f-orbitals.  Herein, by taking single-layer (SL) H-FeBr$_2$ as a representative of the SL H-FeX$_2$ (X=Cl, Br, I) family, we investigated the electronic correlation effects in the magnetic anisotropy and electronic topology of such a system based on first-principles calculations with DFT+\textit{U} approach. Our result is that the magnetic anisotropy energy (MAE) of SL H-FeBr$_2$ shows a non-monotonic evolution behaviour with increasing electronic correlation strength, which is mainly due to the competition between different element-resolved MAEs of Fe and Br. Further investigations show that the evolution of element-resolved MAE arises from the variation of the spin-orbital coupling (SOC) interaction between different orbitals in each atom. Moreover, tuning the strength of the electronic correlation can drive the occurrence of band inversions, causing the system to undergoes multiple topological phase transitions, resulting in a quantum anomalous valley Hall (QAVH) effect. These exotic properties are universal for the SL H-FeX$_2$ (X = Cl, Br, I) family. Our work sheds light on the role of electronic correlation effects in tuning magnetic and electronic structures in the SL H-FeX$_2$ (X = Cl, Br, I) family, which could guide advances in the development of new spintronics and valleytronics devices based on these materials. 

\end{abstract}

\maketitle


\section{\label{sec:level1}INTRODUCTION}
Electrons in materials with open d- or f-shells possess an electronic correlation that would significantly affect the electronic structure in materials and give rise to a variety of exotic properties such as electronic topology\cite{PhysRevB.104.235108,PhysRevB.91.125139,PhysRevLett.121.066402,PhysRevLett.122.016803,PhysRevResearch.3.013265}, magnetism\cite{PhysRevB.97.184404, PhysRevB.95.075124,2021Electron}, and metal-insulator transitions \cite{PhysRevLett.116.116403,PhysRevB.98.075117,PhysRevB.104.035108}. Generally, electron correlation effects are more pronounced in two-dimensional (2D) systems than in three-dimensional (3D) systems with the same chemical composition. This is because Coulomb screening, which suppresses the long-range Coulomb interaction between electrons, is inhibited by dimensionality reduction in 2D systems. Therefore, it is easier to tune the electronic correlation strength of 2D systems experimentally, e.g., through the use of 
structured substrate \cite{2017Coulomb,2019A,PhysRevLett.123.206403} or the growth of multilayer films with different thicknesses\cite{2021Anomalous}. On the other hand, the properties developed by Coulomb engineering have been theoretically revelled in 2D materials\cite{PhysRevB.105.115115,PhysRevB.104.085149,PhysRevB.98.075117,PhysRevB.104.035108}. For example, a coexisting quantum anomalous Hall insulation state in the VSi$_2$P$_4$ monolayer has been proposed with tuning of the Hubbard \textit{U}-constant in first-principles calculations \cite{PhysRevB.104.085149}; By varying the Coulomb repulsion \textit{U}, the location of the metal-insulator transition and the magnetism as well as the superconductivity of some 2D systems were revealed \cite{PhysRevB.98.075117}. Thus, 2D materials provide a good playground to investigate numerous electronic correlation effects.

Recently, a new class of 2D van der Waals systems, single-layer (SL) H-FeX$_2$ (X=Cl, Br, I) \cite{2020The,hu2020concepts,ZHAO202256}, has attracted much attention. It is reported that the SL H-FeCl$_2$ belongs to ferrovalley (FV) materials \cite{2016Concepts} with spontaneously valley polarization coupled with their ferromagnetism. 
 Theoretical calculations have shown that  the orientation of easy magnetization axis of SL H-FeClF would rotate from out-of-plane to in-plane  \cite{PhysRevB.105.104416} with increasing the strength of the electronic correlation (\textit{U$_{\textup{eff}}$}). In addition, the electronic energy bands of SL H-FeClF exhibit a noticeable evolution with \textit{U$_{\textup{eff}}$}, giving rise to topological phase transitions  \cite{PhysRevB.105.104416}.  Intuitively, varying the strength of electronic correlations alters the local distribution of electronic valence charges on the open d-shell of Fe ions and changes the local magnetism and the strength of spin-orbital coupling (SOC). This, of course, drives the evolution of the atomic-scale magnetic anisotropy (MA)  of the system. As we know, the atomic-scale MA is not only tightly related to the performance of the upper limit of the system's magnetic memory, but also in connection with the polarization of valleys in FV materials due to the magneto-valley coupling \cite{PhysRevB.102.035412, PhysRevB.104.085149}. So, it is necessary to study the MA of  SL H-FeX$_2$ (X=Cl, Br, I). Unfortunately, the evolutionary behaviour of MA with \textit{U$_{\textup{eff}}$} of SL H-FeX$_2$ (X=Cl, Br, I) members is unknown. In addition, since the modulation of both the electronic structure and the magnetic properties in each member of the SL H-FeX$_2$ (X=Cl, Br, I) family is all driven by the Coulomb correlation on Fe, the performance in MA of each member of the family would likely share a common trait. On the other hand, different member of SL H-FeX$_2$ (X=Cl, Br, I) has different halogen element, and these different halogen elements should, to some extent, alter the modulation of the electronic and magnetic properties of the system in different manner. The study of these concerns is clearly of great importance in both the scientific interest and the technological importance of spintronics.
 
In this work, we investigated the effect of electronic correlation strength on electronic structures as well as the magnetic anisotropy energy (MAE), represented by SL H-FeBr$_2$. Based on the first-principles calculations with DFT+\textit{U} approach, We found that the trend of MAE with increasing value of \textit{U$_{\textup{eff}}$} imposed on Fe shows a non-monotonic evolution behavior corresponding to a flip between in-plane and out-of-plane magnetization. We suggested that this non-monotonic evolutionary behavior of MAE is  attributed to the competition between the element resolved MAE  contributed by Fe and Br atoms. The above-mentioned MAE behavior with \textit{U$_{\textup{eff}}$} and its underlying mechanism is universal for the SL H-FeX$_2$ (X=Cl, Br, I) family. In addition, the band inversion occurs in SL H-FeBr$_2$ during the rise of \textit{U$_{\textup{eff}}$}, leading to occurrence of topological phase transitions and quantum anomalous valley Hall (QAVH) states. Our work highlights the correlation effects on the MAE and the topological phase transition in the H-FeX$_2$ (X=Cl, Br, I) family.

\section{\label{sec:level1}METHODS}
We performed first-principles density functional theory (DFT) calculations implemented in the Vienna \textit{ab initio} Simulations Package (VASP) \cite{PhysRevB.54.11169} to study all our concerns in the present work. In our theoretical treatment, the plane-wave cut-off energy was set to be 600 eV and the Brillouin zone was sampled with a $12\times 12\times 1$ $\Gamma$-centered $k$-point mesh. The generalized gradient approximation (GGA) with the Perdew-Burke-Ernzerhof (PBE) realization was used for the exchange correlation functional \cite{PhysRevLett.77.3865}.    
To treat the effect of correlation between electrons in the 3d orbitals of Fe atoms, the Dudarev's approach of the Coulomb correction implemented in  DFT+\textit{U} scheme was applied to Fe ions in the system, where only the \textit{U$_{\textup{eff}}$} =$\textit{U}-\textit{J}$ is meaningful (Dudarev approach) \cite{PhysRevB.57.1505}. In addition, the vdW corrections included in the DFT-D3 method was considered \cite{2010A}.
During the structural optimization, both the atomic positions and the lattice constants of each system were fully relaxed until the Hellmann-Feynman forces acting on each atom were less than $10^{-3}$ eV; The electronic convergence criteria was set to be $10^{-6}$eV. Since our concerned system is a two-dimensional nanosheet, a 20\AA vacuum region was added along the direction perpendicular to the surface of the nanosheet was added to avoid the interaction between the periodic images. 
To calculated the Berry curvature, the maximally localized Wannier functions (MLWFs) were constructed using the WANNIER90 package \cite{MOSTOFI20142309}. The edge states were calculated using the iterative Green function method, which is implemented in the WannierTools package \cite{2017WannierTools}.

The MAE is defined as the total energy difference between the in-plane ferromagnetic (FM) configuration (\textit{E$_{\textup{x}}$}) and the out-of-plane FM configuration (\textit{E$_{\textup{z}}$}), namely 
\begin{equation}
MAE = \textit{E$_{\textup{x}}$}-\textit{E$_{\textup{z}}$}. 
\end{equation}

Here, the positive or negative value of MAE indicates that the orientation of easy magnetization axis is along the out-of-plane or in-plane direction. Moreover, the element- and orbital-resolved MAE were calculated from the difference of SOC energies between in-plane and out-of-plane ferromagnetic configurations, i.e., \cite{PhysRevB.101.214436}
\begin{equation}
\Delta{E}_{\textup{SOC}} = E^{x}_{\textup{SOC}}- E^{z}_{\textup{SOC}},
\end{equation}
with 
\begin{equation}
E_{\textup{SOC}} = \langle \frac{h^{2}}{2m^{2}c^{2}}  \frac{1}{r} \frac{dV}{dr} \hat{L} \cdot \hat{S}  \rangle,
\end{equation}
where $V(r)$ is the spherical  part of the effective potential inside the PAW sphere, while $\hat{L}$ and $\hat{S}$ are the orbital and spin angular momentum respectively.  According to the second-order perturbation theory, only about 50\% of the SOC energy difference contributes to the MAE, i.e., $\textup{MAE} \approx \frac{1}{2} \Delta{E}_{\textup{SOC}}$ \cite{ANTROPOV201435,PhysRevB.96.014435}, and the remaining SOC energies might be translated into both crystal-field energy and unquenched orbital moments \cite{2011Is}.  

\section{\label{sec:level1}RESULTS AND DISCUSSION}
\subsection{\label{sec:level2}Basic structural and magnetic properties}

\begin{figure}[ht]
\includegraphics[scale=0.27]{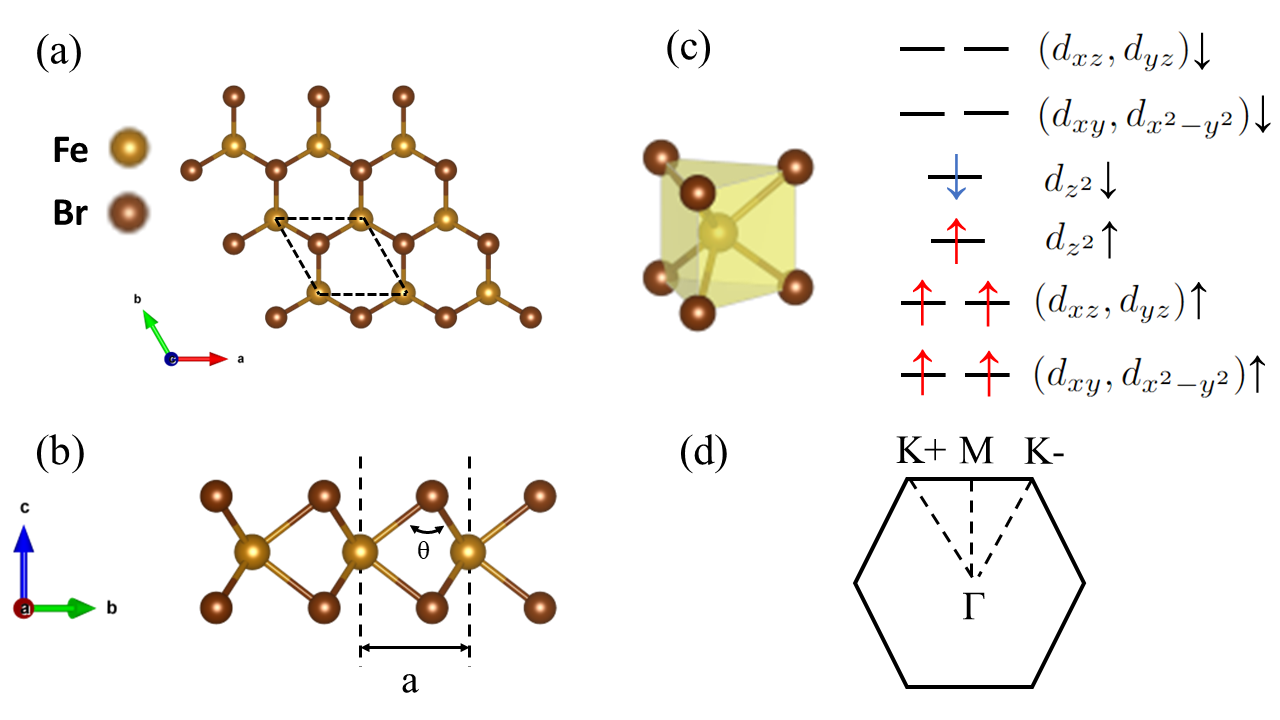}
\caption{
(a) Top and (b) side view of the atomic structure of SL H-FeBr$_2$.(c) The splitting of 3d orbitals of Fe atom under the trigonal prismatic crystal field. The trigonal prismatic crystal structure is also shown. (d)The Brillouin zone with high-symmetry points labeled.
 }
\label{fig1}
\end{figure}

The single layer (SL) H-FeBr$_2$ is a hexagonal structure with the space group of \textit{P$\overline{6}$m}2 (No.187). A monolayer of Fe atoms is sandwiched by two adjacent Br monolayers and each Fe atom is surrounded by six Br atoms, forming a local FeBr$_6$ trigonal prism, as shown in Figs.\ref{fig1}(a) and (b). Our calculations showed that the lattice constant of this system is \textit{a}= 3.57\AA and the bond angle of Fe-Br-Fe is $\theta=85.3^{\circ}$, which are in excellent agreement with the literature \cite{2020The}. In particular, the six Br ions in each FeBr$_6$ trigonal prism provide a local crystal field on each Fe atom (seen in Fig. \ref{fig1}(c)). This causes the 3d orbitals of the Fe atom to split into three groups in the energy landscape: the $d_{z^{2}}$ orbital (denoted as A$_1$), the degenerate $(d_{xy}, d_{x^{2}-y^{2}})$ orbitals (denoted as E$_1$), and the  $(d_{xz}, d_{yz})$ orbitals (denoted as E$_2$), 
 which are schematically displayed in Fig. \ref{fig1}(c). Since the electronic configuration of the Fe$^{2+}$ ion is 3d$^{6}$, the spin-up channels of the 3d orbitals are fully occupied. For the spin-down channel, only the $d_{z^{2}}$ orbital is occupied by an electron, and the other d orbitals are empty. This gives rise to a spin magnetic moment of 4$\mu_{B}$ at each Fe$^{2+}$ ion. 
 
 \begin{figure}[ht]
\includegraphics[scale=0.4]{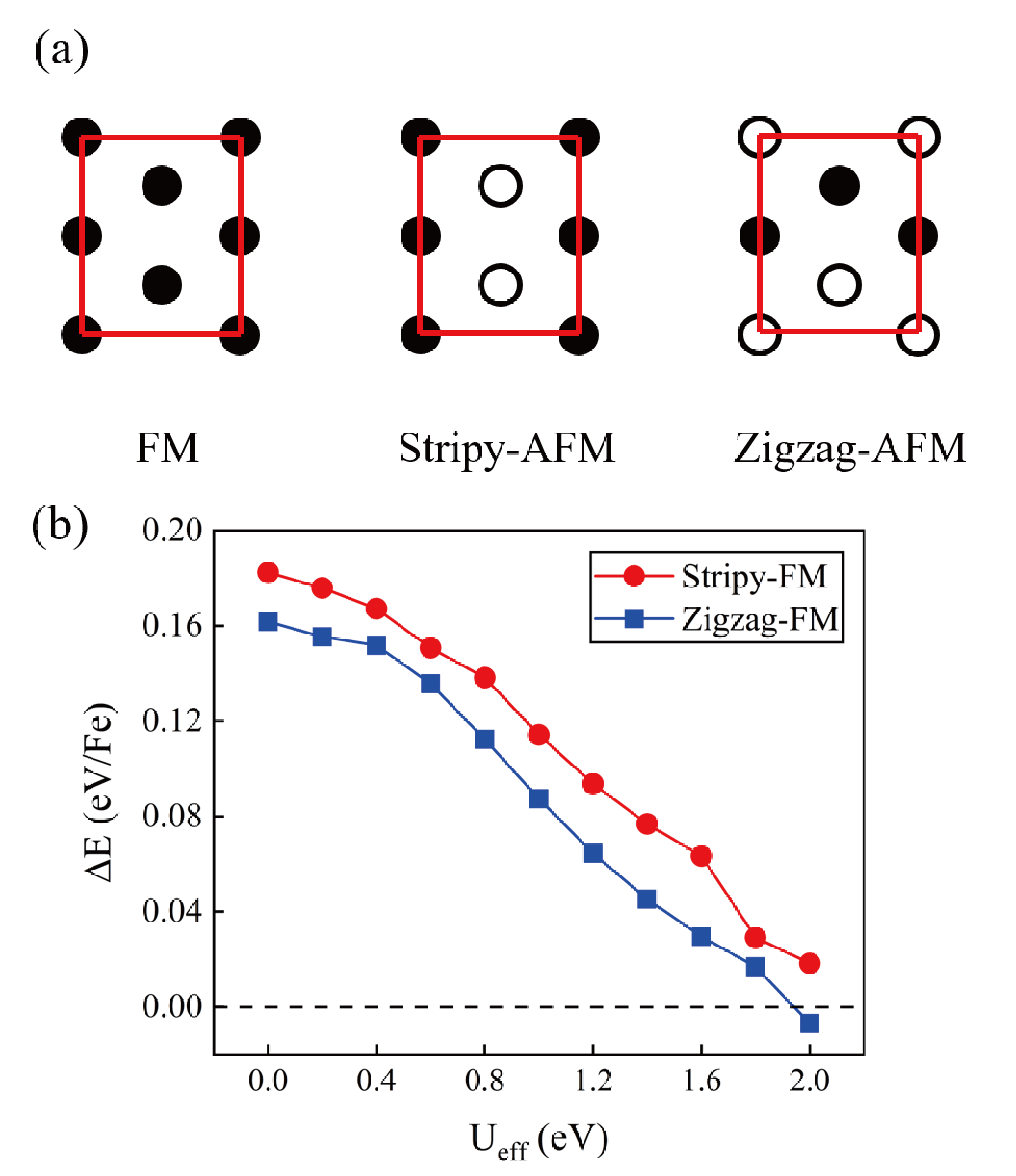}
\caption{
(a) The schematic top views of the FM, Stripy-AFM, and Zigzag-AFM magnetic configurations. The solid and open circles represent spin-up and spin-down states, respectively. 
(b) The evolution of energy difference between FM and Stripy-AFM (Zigzag-AFM) states with \textit{U$_{\textup{eff}}$}. The energy difference is defined as $\Delta E = E_{\textup{Stripy/Zigzag}}-E_{\textup{FM}}$.
 }
\label{fig2}
\end{figure}

To investigate the magnetic ground state of the SL H-FeBr$_2$, three types of magnetic configurations, i.e., a FM configuration, a stripy antiferromagnetic (AFM) configuration and a zigzag AFM configuration (seen in Fig.\ref{fig2} (a)), were considered. Performing calculations of the nanosheet with each concerned magnetic configuration when\textit{U$_{\textup{eff}}$} imposed on Fe and varied from 0.0 to 1.8 eV, we found that the energy of the system with either the stripy AFM or the zigzag AFM is significantly higher than that with the FM configuration (Fig.\ref{fig2}(b)), which agrees well with the previous study \cite{2020The}.


The origin of the above FM state can be understood in terms of the super-exchange interaction.
Note that the bond angle (85.3°) of Fe-Br-Fe in SL H-FeBr$_2$ is close to 90°. Combining this structural feature with the Goodenough-Kanamori-Anderson (GKA) rules \cite{PhysRev.100.564,KANAMORI195987}, we know that the super-exchange interaction between the two nearest neighbouring Fe atoms predominately characterizes the feature of FM interaction. In addition to this FM super-exchange interaction, there is also a weak direct AFM exchange between the two nearest neighbouring Fe ions. However, its strength is usually weaker than that of the FM super-exchange interaction, due to the localization of d-orbitals on each magnetic atom.
Hence, the FM super-exchange interaction dominates the interaction between two nearest neighbouring Fe ions in this nanosheet when \textit{U$_{\textup{eff}}$} is small.

Another aspect shown in Fig. \ref{fig2}(b) is that the energy difference between the stripy AFM and the FM configurations or between the zigzag AFM and the FM configurations (defined as $\Delta E = E_{\textup{Stripy/Zigzag}}-E_{\textup{FM}}$) decreases with increasing the value of \textit{U$_{\textup{eff}}$}, showing a weakening of FM coupling. When \textit{U$_{\textup{eff}}$} reaches 2.0eV, the Zigzag AFM configuration has a lower energy than that of the FM configuration, leading to an AFM ground state.

In fact, this FM-AFM transition behaviour can be understood as a consequence of the competition between the indirect FM superexchange and the direct AFM exchange. As we know, the indirect FM superexchange strength is proportional to 
 $-\frac{ t^{4}_{pd} J^{p}_{H} }{(\Delta_{pd}+U_{d}  )^{4}} $, 
 with $t_{pd}$, $J^{p}_{H}, \Delta_{pd}$ and $U_{d}$ 
representing the hybridization strength between Fe-d orbitals and Br-p orbitals, the Hund's coupling strength of Br-p orbitals, the energy interval between Fe-d orbitals and Br-p orbitals, and the spin-splitting energy of Fe-d orbitals, respectively\cite{PhysRevB.105.085129}. Meanwhile, the strength of direct AFM exchange is proportional to 
$\frac{t_{dd}}{U_{d}}$, 
with $t_{dd}$ being the strength of direct hybridization between d orbitals of nearest neighbouring Fe ions. 

With the increase of \textit{U$_{\textup{eff}}$}, the spin splitting $U_{d}$ increases significantly. From the above formulas, it can be seen that the indirect FM superexchange strength is inversely proportional to the fourth power of the $U_{d}$ value, while the direct AFM exchange strength is only inversely proportional to the first power of the $U_{d}$ value. Obviously, increasing the $U_{d}$ value leads to a more pronounced decrease in the strength of the indirect FM superexchange than that of the direct AFM exchange. As a result, the energy difference between AFM and FM decreases with \textit{U$_{\textup{eff}}$}, giving rise to a transition between AFM and FM magnetic configurations. 

We emphasize that although the energy of stripy AFM state is slightly lower than that of the FM state (for about 0.007eV/Fe) when \textit{U$_{\textup{eff}}$}=2eV, the FM is the most energetically favourable state in most of the considered range (from 0.0eV to 2.0eV) of \textit{U$_{\textup{eff}}$}. Thus, in our following calculations, only the FM state of the system will be investigated.

\subsection{\label{sec:level2}Electronic correlation effects on MAE}

\begin{figure}[t]
\includegraphics[scale=0.27]{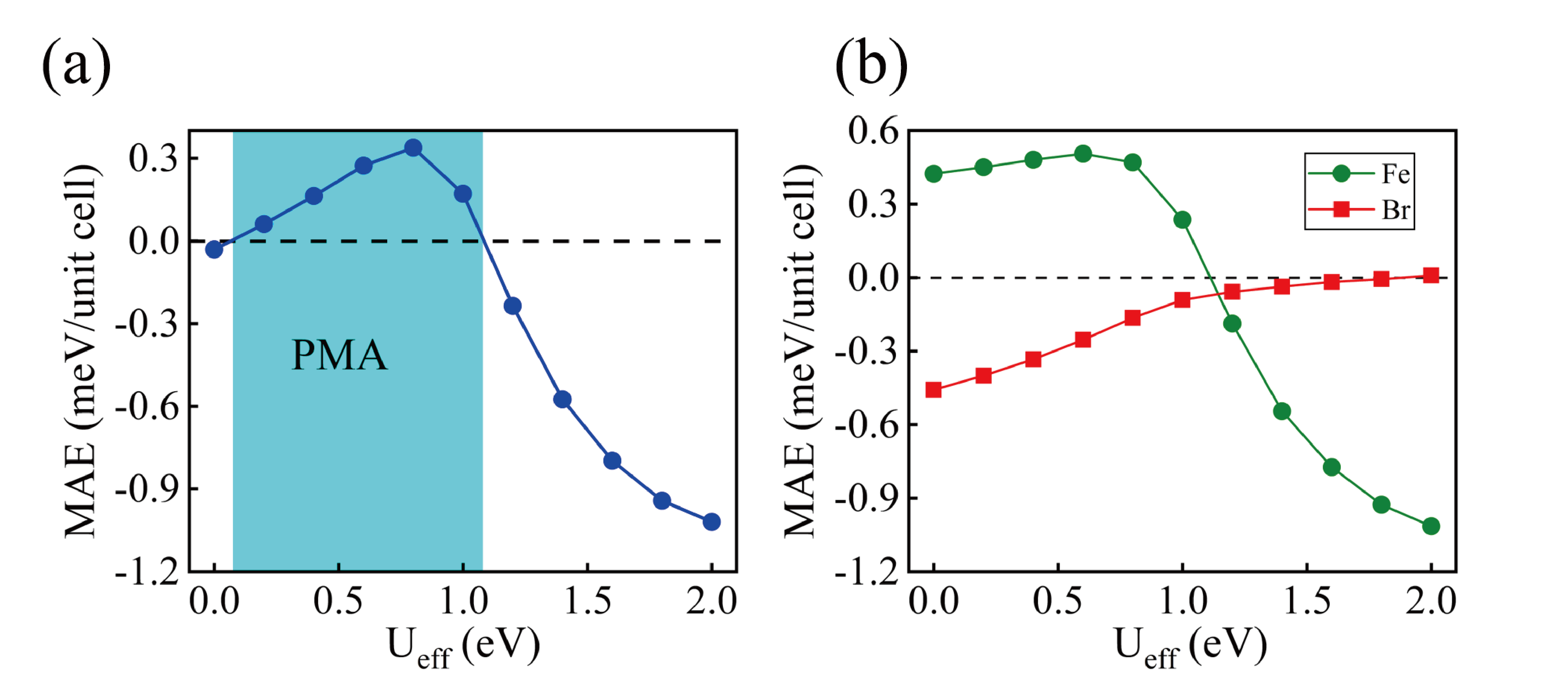}
\caption{
The evolution of (a) total MAE and (b) element-resolved MAE in a unit cell with \textit{U$_{\textup{eff}}$}.
 }
\label{fig4}
\end{figure}
 Now we turn to investigate the evolution of MAE with different correlation strength represented by \textit{U$_{\textup{eff}}$} in Fe. 
Figure \ref{fig4} displays our calculated MAE as a function of \textit{U$_{\textup{eff}}$}. As shown in Fig \ref{fig4}(a), the MAE value increases with increasing the value of \textit{U$_{\textup{eff}}$} until the \textit{U$_{\textup{eff}}$} value reaches 0.8 eV, followed by a rapid decrease in MAE. Therefore, by increasing the value of \textit{U$_{\textup{eff}}$}, the MAE varies from negative values to positive values and then to negative values again. This corresponds to the change of the magnetic state from the out-of-plane FM state to the in-plane FM state before returning to the out-of-plane FM state. The system prefers an in-plane FM state if $\textit{U$_{\textup{eff}}$}$\textless0.1eV or $\textit{U$_{\textup{eff}}$}$\textgreater1.1 eV, otherwise, the systems shows an out-of-plane FM state. So, there exists a perpendicular magnetic anisotropy (PMA) in the range of 0.1 eV\textless\textit{U$_{\textup{eff}}$}\textless1.1 eV.  
   
Basically, the entire MAE of a system is contributed from each atom. Thus, the element-resolved MAE as a function of \textit{U$_{\textup{eff}}$} is calculated, which is shown in Fig. \ref{fig4}(b). It is remarkable that the MAE from Fe and Br elements show a completely different behavior with increasing \textit{U$_{\textup{eff}}$}: the MAE of Br element (denoted as Br-MAE) increases monotonically with increasing \textit{U$_{\textup{eff}}$}; However, the MAE of Fe (denoted as Fe-MAE) remains almost constant before \textit{U$_{\textup{eff}}$} reaches 0.8 eV, and when \textit{U$_{\textup{eff}}$}\textgreater0.8 eV Fe-MAE decreases sharply and varies from positive values to negative values. 
Apparently, the non-monotonic behaviour of the whole MAE in a unit cell could be the result of the competition of the element-resolved MAE between Fe and Br. For \textit{U$_{\textup{eff}}$}\textless0.8 eV, the enhancement of Br-MAE dominates, which is responsible for the increase in the total MAE. For \textit{U$_{\textup{eff}}$}\textgreater0.8 eV, the faster change of the Fe-MAE dominates and gives rise to a decrease in the total MAE. 

\begin{figure*}[ht]
\includegraphics[scale=0.5]{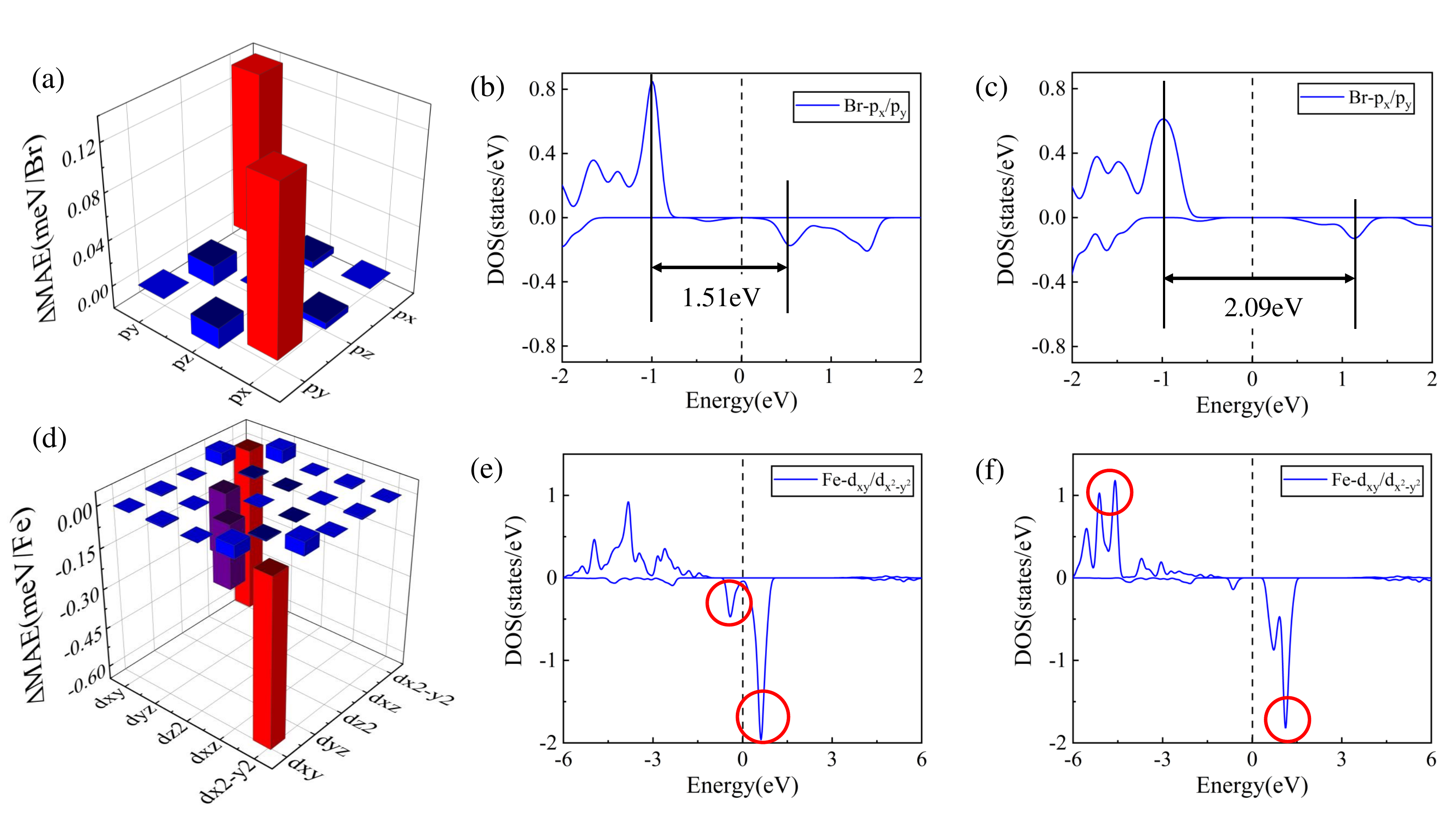}
\caption{
(a) The difference of orbital-resolved MAE of each Br atom between \textit{U$_{\textup{eff}}$}=0.0 eV and \textit{U$_{\textup{eff}}$}=2.0 eV, which is defined as $\Delta\textup{MAE} = {\textup{MAE}}_{\textit{U$_{\textup{eff}}$}= 2 eV}-{\textup{MAE}}_{\textit{U$_{\textup{eff}}$}= 0 eV}$. The spin-polarized DOS of Br-$p_{x}$ and Br-$p_{y}$ orbitals at (b) \textit{U$_{\textup{eff}}$}=0.0eV and (c) \textit{U$_{\textup{eff}}$}=2.0 eV is given. 
(d) The difference of orbital resolved MAE of each Fe atom between \textit{U$_{\textup{eff}}$}=0.8eV and \textit{U$_{\textup{eff}}$}=2.0 eV, which is defined as $\Delta\textup{MAE} = {\textup{MAE}}_{\textit{U$_{\textup{eff}}$}=2 eV}-{\textup{MAE}}_{\textit{U$_{\textup{eff}}$}=0.8 eV}$. The spin-polarized DOS of Fe-d$_{xy}$ and Fe -$d_{x^{2}-y^{2}}$ orbitals at (e) \textit{U$_{\textup{eff}}$}=0.8 eV and (f) \textit{U$_{\textup{eff}}$}=2.0 eV is given. 
}
\label{fig5}
\end{figure*}
    
Why does the evolution of Br-MAE behave differently than that of Fe-MAE with changing \textit{U$_{\textup{eff}}$}?  To uncover the nature underlying this concern, we recall that based on the second-order perturbation theory, the MAE essentially correlates with the SOC interaction between the occupied and unoccupied states around the Fermi level, which is expressed as \cite{PhysRevB.47.14932,PhysRevB.103.195402}
\begin{equation}
\begin{aligned}
& \textup{MAE} \\
  & =  \sum_{\sigma,\sigma'}(2\delta_{\sigma,\sigma'}-1)\xi^{2}\sum_{o^{\sigma},u^{\sigma'}} \frac{|\langle o^{\sigma} |\hat{L}_{z} |u^{\sigma'} \rangle |^{2} - |\langle o^{\sigma} |\hat{L}_{x} |u^{\sigma'} \rangle |^{2}}{E^{\sigma'}_{u}-E^{\sigma}_{o}}    
\end{aligned}
\end{equation}

Here $\xi$ is the strength of SOC, $\sigma$ and $\sigma'$ are spin indexs. $\hat{L}_{z}$ and $\hat{L}_{x}$ are angular momentum operators.   $|o^{\sigma} \rangle $ ($|u^{\sigma'} \rangle $) is the occupied (unoccupied) state with spin $\sigma$ ($\sigma'$), whose energy is $E^{\sigma}_{o}$ ($E^{\sigma'}_{u}$).  According to the expression above,
the MAE is sensitive to the energy interval between occupied and unoccupied states, $ E^{\sigma'}_{u}-E^{\sigma}_{o}$. Thus, by tuning the value of \textit{U$_{\textup{eff}}$}, the value of $ E^{\sigma'}_{u}-E^{\sigma}_{o}$ could change, giving rise to the variation of MAE.

Furthermore, we computed the orbital-resolved MAE for each Br atom with \textit{U$_{\textup{eff}}$}= 0 eV and 2 eV  and for each Fe atom with \textit{U$_{\textup{eff}}$}=0.8 eV and 2 eV respectively (as shown in Fig. S1 in the Supplemental Material \cite{SM}), followed by making difference $\Delta \textup{MAE} = {\textup{MAE}}_{ \textit{U$_{\textup{eff}}$}= 2 eV}-{\textup{MAE}}_{ \textit{U$_{\textup{eff}}$}= 0 eV}$ for Br and $\Delta \textup{MAE} = {\textup{MAE}}_{\textit{U$_{\textup{eff}}$}=2 eV}-{\textup{MAE}}_{\textit{U$_{\textup{eff}}$}=0.8 eV}$ for Fe, as displayed in Fig.\ref{fig5}(a) and \ref{fig5}(d). 

In the case of Br, it is observed in Fig. \ref{fig5}(a) that as the value of  \textit{U$_{\textup{eff}}$} increases from 0.0 eV to 2.0 eV, the $p_{x}-p_{y}$ orbital pairs make more positive contributions to the Br-MAE, while those from other orbital pairs have only minor contributions, thus only the Br $p_{x}$ and $p_{y}$ orbitals are considered. To understand the phenomenon observed above, we calculated the density of states (DOS) of spin-polarized Br-$p_{x}$ and Br-$p_{y}$ orbitals in SL H-FeBr$_{2}$ when \textit{U$_{\textup{eff}}$} = 0.0 eV and \textit{U$_{\textup{eff}}$} = 2.0 eV, which are shown in Fig. \ref{fig5}(b) and \ref{fig5}(c). It can be observed that the $p_{x}$ and $p_{y}$ orbitals of Br are all degenerate. Furthermore, the unoccupied states of Br near the Fermi level are the spin-down $p_{x/y}$ states and the occupied states of Br near the Fermi level are the spin-up $p_{x/y}$ states.

We first investigate the MAE contributed by the SOC interaction between the unoccupied spin-down Br-$p_{y}$ states (denoted as $|p^{\downarrow}_{y,u} \rangle$) and the occupied spin-up Br-$p_{x}$ states (denoted as $|p^{\uparrow}_{x,o}\rangle$), which can be expressed as:

\begin{equation}
\begin{aligned}
    \textup{MAE}_{\textup{Br-$p_{x/y}$}} & =  -\xi ^{2} \sum_{o,u} \frac{|\langle p^{\uparrow}_{x,o}|\hat{L_{z}}| p^{\downarrow}_{y,u} \rangle |^{2} -|\langle p^{\uparrow}_{x,o}|\hat{L_{x}}| p^{\downarrow}_{y,u} \rangle |^{2}}{E^{\downarrow}_{p_{y,u}}  - E^{\uparrow}_{p_{x,o}} }.\\
\end{aligned}
\end{equation}

Noting that the matrix elements are \cite{doi:10.1021/acs.inorgchem.9b00687}
\begin{equation}
    \langle p_{x} |\hat{L_{z}} | p_{y} \rangle = i, 
\end{equation}
and
\begin{equation}
    \langle p_{x} |\hat{L_{x}} | p_{y} \rangle = 0.
\end{equation}
 Thus, the $\textup{MAE}_{\textup{Br-$p_{x/y}$}}$  can be simplified as :
 \begin{equation}
 \begin{aligned}
     \textup{MAE}_{\textup{Br-$p_{x/y}$}} = -\xi ^{2} \sum_{o,u} \frac{1}{E^{\downarrow}_{p_{y,u}}  - E^{\uparrow}_{p_{x,o}}}.
 \end{aligned}
 \end{equation}
 
Apparently, the $\textup{MAE}_{\textup{Br-$p_{x/y}$}}$ contributes to the negative value of Br-MAE. Meanwhile, the amplitude of the $\textup{MAE}_{\textup{Br-$p_{x/y}$}}$ is inversely proportional to the energy gap between the occupied spin-up $p_{x}$ states and the unoccupied spin-down $p_{y}$ states near the Fermi level. Here, the energy gap is approximately described as the energy difference between the two major peaks in DOS of occupied spin-up $p_{x}$ states and unoccupied spin-down $p_{y}$ states. Adjusting the value of \textit{U$_{\textup{eff}}$} from 0 eV to 2 eV, the aforementioned energy gap  was changed from 1.51 eV to 2.09 eV, as shown in Fig. \ref{fig5}(b) and (c). This increase in the energy gap gives rise to a smaller magnitude of negative $\textup{MAE}_{\textup{Br-$p_{x/y}$}}$, which is responsible for decreasing the value of negative Br-MAE. 

Recall that the $p_{x}$ and $p_{y}$ orbitals of Br are degenerate, by doing the same analysis on the unoccupied spin-down $p_{x}$ states and the occupied spin-up $p_{y}$ states of Br near the Fermi level, we found that the gap width between the main peaks in the DOS corresponding to these two orbitals increase as well, resulting in a decreasing magnitude in the negative Br-MAE. Therefore, the negative Br-MAE contributes less with an increase in \textit{U$_{\textup{eff}}$}.

For Fe, it is shown in Fig. \ref{fig5}(d) that as the value of  \textit{U$_{\textup{eff}}$} increases from 0.8 eV to 2.0 eV, the $d_{xy}-d_{x^{2}-y^{2}}$ orbital pairs make more negative contributions to the Fe-MAE, and those from other orbital pairs have minor contributions, thus only $d_{x^{2}-y^{2}}$ and $d_{xy}$ orbitals of Fe are considered. To explain this phenomenon, the spin-polarized DOS of Fe-$d_{x^{2}-y^{2}}$ orbital and Fe-$d_{xy}$ orbital in SL H-FeBr$_2$ are calculated when \textit{U$_{\textup{eff}}$} =0.8 eV and 2.0 eV, respectively, which are shown in Figs. \ref{fig5}(e) and (f)). 
It is easy to find that the $d_{xy}$ and $d_{x^{2}-y^{2}}$ orbitals of Fe are all degenerate.

In the case of \textit{U$_{\textup{eff}}$} =0.8 eV, the occupied spin-down $d_{x^{2}-y^{2}}$ states (denoted as $|d^{\downarrow}_{x^{2}-y^{2},o}\rangle $) interact strongly with the unoccupied spin-down $d_{xy}$ states (denoted as $|d^{\downarrow}_{xy,u} \rangle$) near the Fermi level under SOC (as circled in Fig.\ref{fig5} (e)),
It is noted that the orientations of the spins in these states are same. The MAE of Fe contributed by these two states is called as spin-conserved MAE, denoted as $\textup{MAE}_{\textup{Fe},\textup{spin-conserved}}$. 
Here, we have
\begin{equation}
\begin{aligned}
    &\textup{MAE}_{\textup{Fe},\textup{spin-conserved}} \\
    & =  \xi ^{2} \sum_{o,u} \frac{|\langle d^{\downarrow}_{xy,u}|\hat{L_{z}}| d^{\downarrow}_{x^{2}-y^{2},o} \rangle |^{2} -|\langle d^{\downarrow}_{xy,u}|\hat{L_{x}}| d^{\downarrow}_{x^{2}-y^{2},o} \rangle |^{2}}{E^{\downarrow}_{d_{xy,u}}  - E^{\downarrow}_{d_{x^{2}-y^{2},o}} }\\
\end{aligned}
\end{equation}
with using the matrix elements  \cite{doi:10.1021/acs.inorgchem.9b00687} 
\begin{equation}
    \langle d_{xy} |\hat{L_{z}} | d_{x^{2}-y^{2}} \rangle = 2i, 
\end{equation}
and
\begin{equation}
    \langle d_{xy} |\hat{L_{x}} | d_{x^{2}-y^{2}} \rangle=0,
\end{equation}

the $\textup{MAE}_{\textup{Fe},\textup{spin-conserved}}$ can be simplified as
\begin{equation}
\begin{aligned}
    \textup{MAE}_{\textup{Fe},\textup{spin-conserved}} = \xi ^{2} \sum_{o,u} \frac{4}{E^{\downarrow}_{d_{xy,u}}  - E^{\downarrow}_{d_{x^{2}-y^{2},o}}}
\end{aligned}    
\end{equation}

Clearly, the value of $\textup{MAE}_{\textup{Fe},\textup{spin-conserved}}$ is positive, which contributes to the positive Fe-MAE when \textit{U$_{\textup{eff}}$} is about 0.8 eV. 
However, when \textit{U$_{\textup{eff}}$} reaches 2 eV, the occupied spin-down $d_{x^{2}-y^{2}}$ orbitals almost vanish on the VBM, as seen in Fig. \ref{fig5}(f). This means that the SOC interaction between unoccupied spin-down $d_{xy}$ states and occupied spin-down $d_{x^{2}-y^{2}}$ states  almost disappears at this value of \textit{U$_{\textup{eff}}$}. 
Instead, the unoccupied spin-down $d_{xy}$ states near the Fermi level mainly interact with the occupied spin-up $d_{x^{2}-y^{2}}$ states (denoted as $|d^{\uparrow}_{x^{2}-y^{2},o}\rangle  $) lying deeply in the valance bands, as circled in Fig. \ref{fig5}(f). This SOC interaction between these two states with anti-paralleled spins contributes to the Fe-MAE as well, and we denoted this MAE contribution as $\textup{MAE}_{\textup{Fe},\textup{spin-flip}}$. It can be written as:
\begin{equation}
\begin{aligned}
        &\textup{MAE}_{\textup{Fe},\textup{spin-flip}} \\
    & =  -\xi ^{2} \sum_{o,u} \frac{|\langle d^{\downarrow}_{xy,u}|\hat{L_{z}}| d^{\uparrow}_{x^{2}-y^{2},o} \rangle |^{2} -|\langle d^{\downarrow}_{xy,u}|\hat{L_{x}}| d^{\uparrow}_{x^{2}-y^{2},o} \rangle |^{2}}{E^{\downarrow}_{d_{xy,u}}  - E^{\uparrow}_{d_{x^{2}-y^{2},o}} }\\
    & = -\xi ^{2} \sum_{o,u} \frac{4}{E^{\downarrow}_{d_{xy,u}}  - E^{\uparrow}_{d_{x^{2}-y^{2},o}}}
\end{aligned}
\end{equation}

\begin{figure*}[ht]
\includegraphics[scale=0.42]{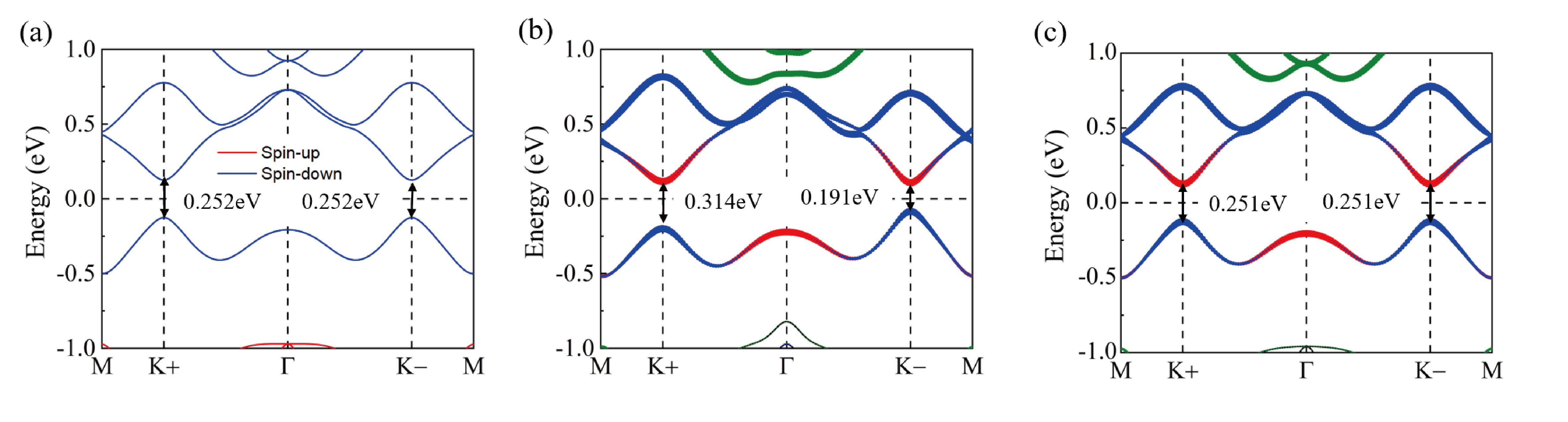}
\caption{
The band structure of ferromagnetic SL H-FeBr$_2$ calculated with (a) spin-polarized case without including SOC, (b) out-of-plane magnetism with SOC, and (c) in-plane magnetization magnetism with SOC. In (b) and (c), the red, blue and green dots represent Fe-A$_{1}$, Fe-E$_{1}$, and Fe-E$_{2}$ orbitals, respectively. 
 }
\label{fig3}
\end{figure*}

The value of $\textup{MAE}_{\textup{Fe},\textup{spin-flip}}$ is negative, which is now responsible for the negative Fe-MAE. 
Note that the $d_{xy}$ and $d_{x^{2}-y^{2}}$ orbitals of Fe are degenerate in this case, we can do the same analysis on the occupied spin-up $d_{xy}$ states and unoccupied spin-down $d_{x^{2}-y^{2}}$ states of Fe near the Fermi level. After carefully examining the features of the DOS corresponding to these two orbitals, we found that the gap between the major peaks in the DOS becomes narrow with increasing \textit{U$_{\textup{eff}}$}. This directly contributes to the negative Fe-MAE value. In total, with increasing \textit{U$_{\textup{eff}}$}, the competition between $\textup{MAE}_{\textup{Fe},\textup{spin-conserved}}$ and $\textup{MAE}_{\textup{Fe},\textup{spin-flip}}$ causes the Fe-MAE to switch from positive values to negative values .

\subsection{\label{sec:level2}Evolution of band structures with electronic correlation strength}

\begin{figure*}[ht]
\includegraphics[scale=0.25]{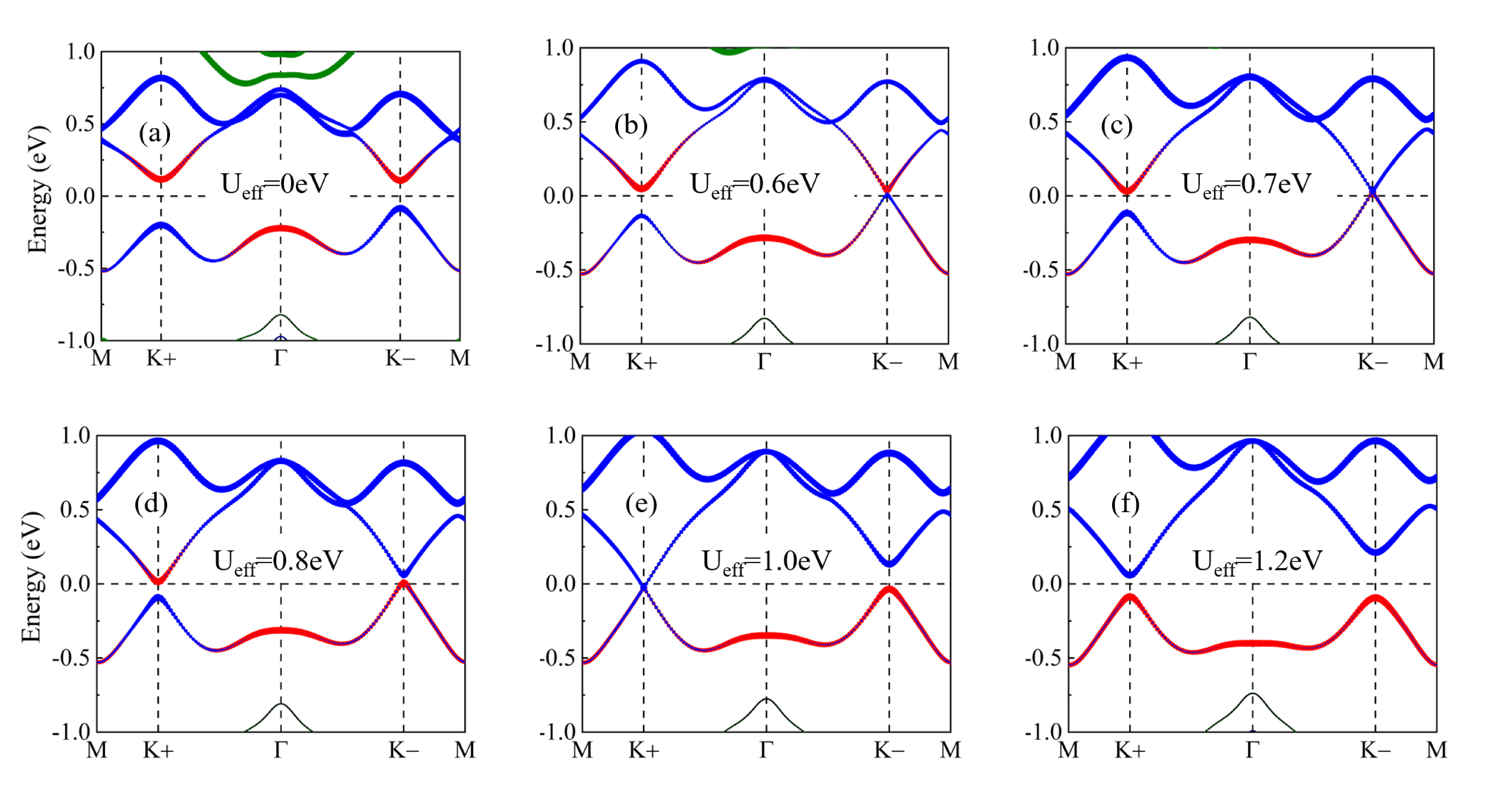}
\caption{
The band structure of SL H-FeBr$_2$ calculated at (a) \textit{U$_{\textup{eff}}$}=0.0 eV, (b) \textit{U$_{\textup{eff}}$}=0.6 eV, (c) \textit{U$_{\textup{eff}}$}=0.7 eV, (d) \textit{U$_{\textup{eff}}$}=0.8 eV, (e) 
\textit{U$_{\textup{eff}}$}=1.0 eV, (f)
\textit{U$_{\textup{eff}}$}=1.2 eV. The red, blue and green dots represent Fe-A$_{1}$, Fe-E$_{1}$, and Fe-E$_{2}$ orbitals, respectively. 
}
\label{fig6}
\end{figure*}

\begin{figure*}[ht]
\includegraphics[scale=0.65]{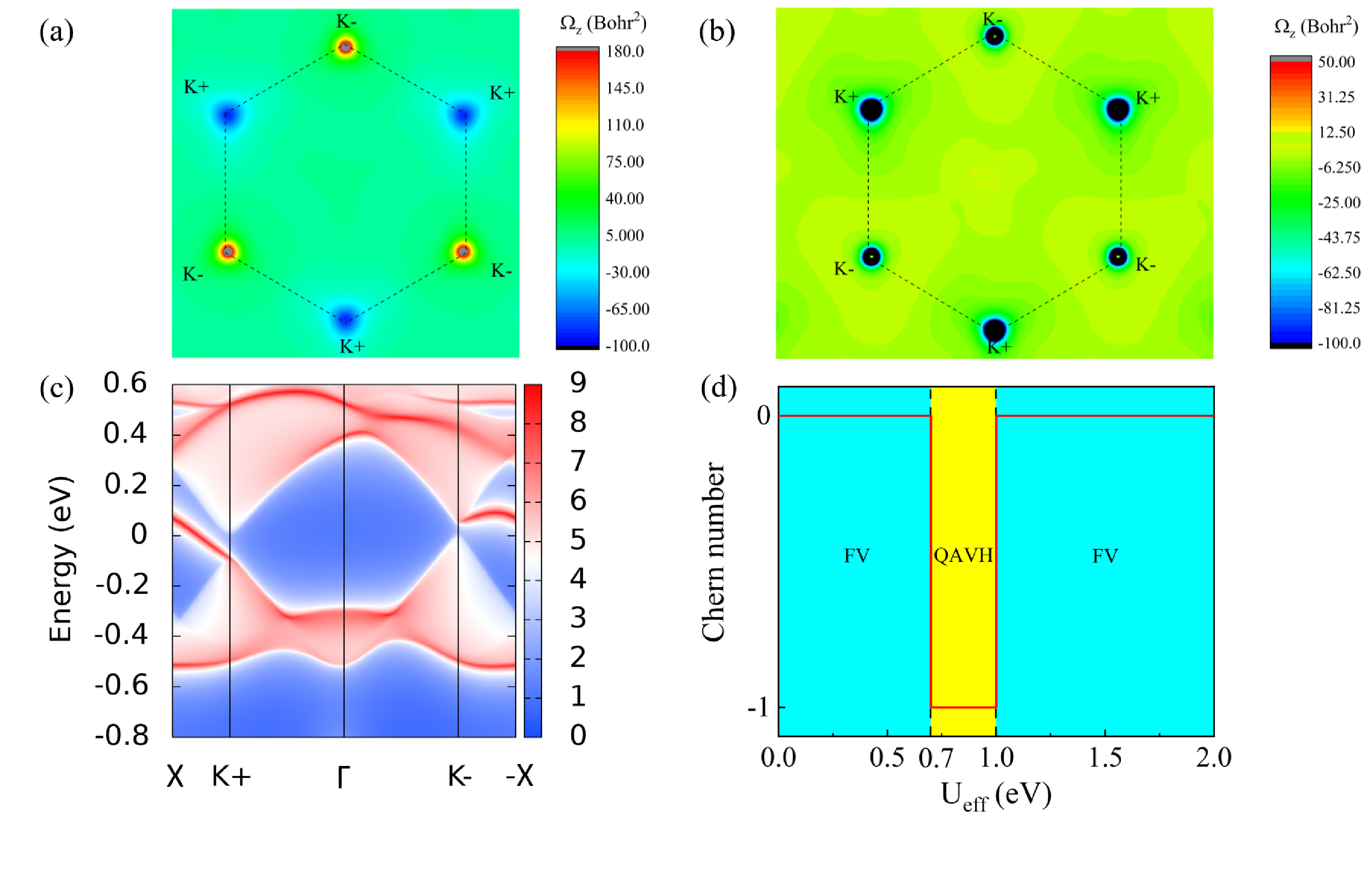}
\caption{
The Berry curvature of SL H-FeBr$_2$ in the 2D Brillouin zone calculated at (a) \textit{U$_{\textup{eff}}$}=0eV and (b) \textit{U$_{\textup{eff}}$}=0.8 eV;
Both units of Berry curvature in (a) and (b) are Bohr$^{2}$.
(c) The topological edge state of SL H-FeBr$_2$ along the (100) direction calculated at \textit{U$_{\textup{eff}}$}=0.8 eV;
(d) The Topological phase diagram of SL H-FeBr$_2$ with varied \textit{U$_{\textup{eff}}$}. 
}
\label{fig7}
\end{figure*}

Let us again pay our attention to Figs. \ref{fig5}(e) and (f). It can be observed that the electronic structures associated with the Fe-3d states vary greatly with \textit{U$_{\textup{eff}}$}, which could be reflected in the variation in the energy bands. Therefore, we carefully calculated the band structures with different \textit{U$_{\textup{eff}}$} values.

For comparison, the band structure of FM SL H-FeBr$_{2}$ at \textit{U$_{\textup{eff}}$} = 0 eV was first examined. In this case, the spin-polarized band structure without considering the SOC is plotted in Fig. \ref{fig3}(a). It can be clearly seen that the spin-up and spin-down channels are split, where the spin-down component dominates near the Fermi level.
Meanwhile, the electronic band structure with spin-down has a band gap of about 0.253 eV. Therefore, this system characterizes the FM half-semiconductor. In addition, the energy values of the valence band maximum (VBM) at K+ and K- are equal. This scenario also occurs at the conduction band minimum (CBM) at the two k-points. Thus, there are degenerate valleys in this energy band. This feature indicates that SL H-FeBr$_{2}$ belongs to the 2D valleytronic materials \textbf{\cite{2016Valleytronics}}.

In fact, the SOC cannot be ignored in this system, so the SOC is taken into account in the following treatment. As mentioned before, the easy axis of magnetization of this system can be in the \textit{ab} plane or the out-of-plane. When the easy axis of magnetization is in the \textit{ab} plane, the calculated energy bands show that the energy gap at K+ and K- are equal (Fig. \ref{fig3}(c)), and since the CBM (VBM) at K+ and K- are degenerate, there is no spontaneous polarization of valleys. When we switch the easy axis to out-of-plane, the band gap width at the K+ point is 0.314 eV and that at the K- point is 0.191 eV (Figure \ref{fig3}(b)), showing the spontaneous polarization of the valley. In this case the system is in the FV state. Strikingly, the difference of the band gap between K+ and K- is 123 meV, being larger than other typical FV materials such as SL H-FeCl$_{2}$ (106 meV) \cite{hu2020concepts}, Nb$_{3}$I$_{8}$ (107 meV) \cite{PhysRevB.102.035412}, LaBr$_{2}$ (33 meV) \cite{2019Single} and MnPX$_{3}$ (43 meV) \cite{doi:10.1073/pnas.1219420110}. Apparently, the valley state strongly couples with the magnetization direction in SL H-FeBr$_{2}$, which is explained theoretically in the Appendix \ref{app}. According to the previous reports \cite{PhysRevB.105.104416,2020The}, such FV materials with out-of-plane magnetization potentially possess topological nontrivial states. Therefore, in our following calculations, the magnetization orientation was set to be out-of-plane.

The evolution of electronic band structures with \textit{U$_{\textup{eff}}$} is investigated, and the representative band structures with different \textit{U$_{\textup{eff}}$} values are shown in Fig. \ref{fig6}. When \textit{U$_{\textup{eff}}$} increases to 0.6 eV, both the conduction bands and valence bands move towards the Fermi level, thereby reducing the band gap at the K+ and K- points. When increasing the value of \textit{U$_{\textup{eff}}$} to 0.7 eV, the band gap at the K+ point is still non-zero, but the band gap at the K- point is closed (as shown in Fig. \ref{fig6}(c)). This shows that the system has Dirac-like band crossing characteristic at K- \cite{RevModPhys.90.015001}. In this case, the system becomes the so-called half valley metal (HVM) \cite{hu2020concepts}, which can provide a massless elementary excitation that potentially contributes to well-behaved charge transport. 

The band gap closed at the K- point reopens when \textit{U$_{\textup{eff}}$} is slightly larger than 0.7 eV . 
Interestingly, at the K- point, the low-lying E$_{1}$ orbitals on the VBM shift to the CBM, while the high-lying A$_{1}$ orbital on the CBM shifts down to the VBM (as shown in Fig. \ref{fig6}(d)) . Despite of this, the orbital composition of the energy band at the K+ point remains intact, forming a single-valley band-inverted state. When the value of \textit{U$_{\textup{eff}}$} is adjusted to about 1.0 eV, the bandgap at the K- point reopens, and the bandgap at the K+ point closes (as shown in Fig. \ref{fig6}(e)). Compared to the case of \textit{U$_{\textup{eff}}$}=0.8 eV, the orbital composition of CBM and VBM at the K+ point are reversed here. Meanwhile, we found that when \textit{U$_{\textup{eff}}$} is larger than 1.0 eV, the conduction band (valance band) near the Fermi level is contributed only by the orbitals of Fe-E$_{1}$ (A$_{1}$) (as shown in Fig. \ref{fig6}(f)). If the value of \textit{U$_{\textup{eff}}$} is further increased, Fe-A$_{1}$ orbitals and Fe-E$_{1}$ orbitals are no longer entangled with each other on the conduction band (valance band) near the Fermi level. During this disentanglement process with increasing \textit{U$_{\textup{eff}}$}, the SOC interaction between the occupied spin-down Fe-d$_{xy}$ (d$_{x^{2}-y^{2}}$) on the valance bands near the Fermi level and the unoccupied spin-down Fe-d$_{x^{2}-y^{2}}$ (d$_{xy}$) orbitals on the conduction bands near the Fermi level weakens, which is the origin of the rapid  switch in Fe-MAE. 


To further characterize the effect of \textit{U$_{\textup{eff}}$} on valley-contrasting physics, the Berry curvature along $\textit{z}$ direction was calculated based on the Kubo formula \cite{PhysRevLett.49.405}:
\begin{equation}
    \Omega_{z} (\textbf{k}) = -\sum_{n}\sum_{n\neq n'} f_{n} \frac{2 \textup{Im} \langle \psi_{n\textbf{k}} | v_{x} | \psi_{n'\textbf{k}} \rangle \langle \psi_{n'\textbf{k}} | v_{y} | \psi_{n\textbf{k}} \rangle }{(E_{n}-E_{n'})^{2}}
\end{equation}
Here, $f_{n}$ is the Fermi-Dirac distribution function, $E_{n}$ is the eigenvalue of Bloch state $| \psi_{n\textbf{k}} \rangle $, and $v_{x/y}$ is the vector operator.

The k-resolved Berry curvatures for \textit{U$_{\textup{eff}}$} = 0 eV and \textit{U$_{\textup{eff}}$} = 0.8 eV are plotted in Fig. \ref{fig7}(a) and Fig. \ref{fig7}(b), respectively. In the case of \textit{U$_{\textup{eff}}$} = 0 eV, the Berry curvatures at K+ and K- have peaks of opposite signs and different absolute values, which originate from the spontaneous breaking of time-reversal symmetry and space-reversal symmetry. For such a system with a single valley band inversion, the Berry curvature peaks at K+ and K- have the same sign when \textit{U$_{\textup{eff}}$} = 0.8 eV. In this case, the full-space integral over the Berry curvature yields a nonzero Chern number, which shows a topologically nontrivial character. To uncover the nature of this topologically nontrivial feature, we computed the edge states along the (100) direction in the system at \textit{U$_{\textup{eff}}$} = 0.8 eV. As shown in Fig. \ref{fig7}(c), there is a gapless chiral edge state, which simultaneously connecting the valence band at K+ and the conduction band at K-. Such gapless chiral edge states are the fingerprints of quantum anomalous Hall (QAH) states in topologically nontrivial systems with ferromagnetism \cite{doi:10.1146/annurev-conmatphys-031115-011417,doi:10.1146/annurev-conmatphys-033117-054144}. Furthermore, the anomalous Hall conductivity (AHC) is calculated by the following formula \cite{PhysRevLett.88.207208,PhysRevLett.92.037204}:
\begin{equation}
\sigma_{xy} = \frac{e^{2}}{h} \int_{BZ} \frac{d\textbf{k} }{(2\pi)^{2}} \Omega_{z} (\textbf{k})   
\end{equation}
 As shown in Fig. S2 in the Supplemental Material \cite{SM}), the AHC in the energy gap is  $-\frac{e^{2}}{h}$ , confirming a QAH state with a Chern number of \textit{C} = -1 in this system. Note that this single-valley band-inverted state with QAH effect is the so-called QAVH state \cite{hu2020concepts}. Therefore, when the electron correlation strength is adequate, SL H-FeBr$_{2}$ can possess a QAVH state, which has great potential applications for the development of spintronic devices. 

In principle, the presence of the QAVH state in the system is relevant to the value of \textit{U$_{\textup{eff}}$}. We therefore carefully look for the range of \textit{U$_{\textup{eff}}$} in which the QAVH effect appears in the system. Fig. \ref{fig7}(d) displays the topological phase diagram in which the QAVH state is predicted to exist in the range of \textit{U$_{\textup{eff}}$} varying from 0.7 eV to 1.0 eV. As we know, the QAVH state can only survive when the system hosts PMA. The value of \textit{U$_{\textup{eff}}$} corresponding to the QAVH state ranges from 0.7 eV to 1.0 eV, which belongs to the \textit{U$_{\textup{eff}}$} value interval of PMA (from 0.1eV to 1.1eV). Thus, the QAVH state could naturally exist in this system without external magnetic fields. 

\subsection{\label{sec:level2}Discussions}
 \begin{figure*}[ht]
\includegraphics[scale=0.25]{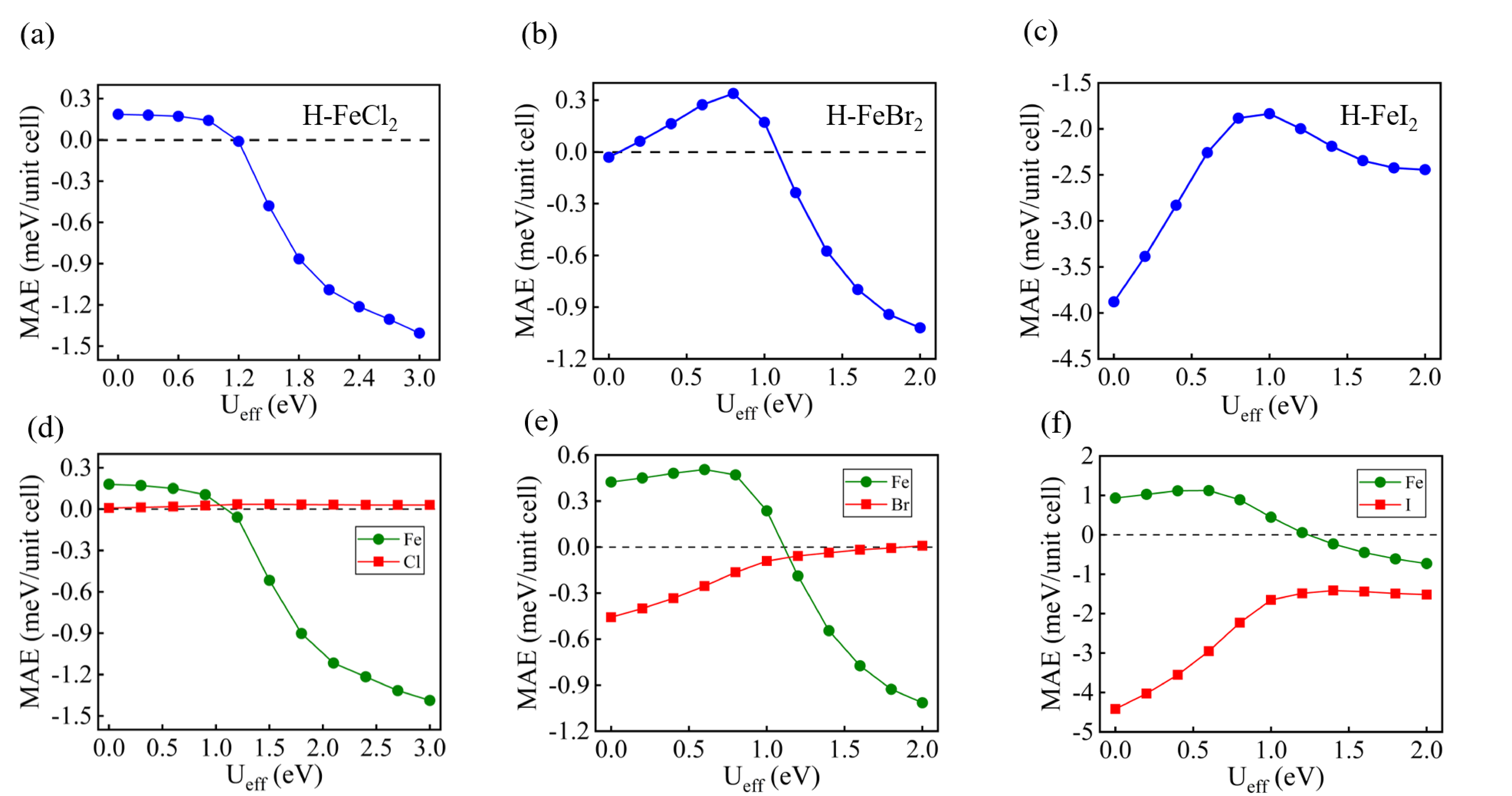}
\caption{
The evolution of total MAE of (a) SL H-FeCl$_2$, (b) SL H-FeBr$_2$, (c) SL H-FeI$_2$. The evolution of element-resolved MAE of (d) SL H-FeCl$_2$, (e) SL H-FeBr$_2$, (f) SL H-FeI$_2$ are also given.  
}
\label{fig8}
\end{figure*}

We are now curious if the behavior of \textit{U$_{\textup{eff}}$}-dependent MAE in SL H-FeBr$_2$ is present in SL H-FeX$_2$ (X = Cl, I), which are the counterparts of SL H-FeBr$_2$. To this end, the total MAEs of SL H-FeX$_2$ (X = Cl, I) in FM state were computed, which are all plotted in Figs. \ref{fig8}(a)-(c). Obviously, as the value of \textit{U$_{\textup{eff}}$} increases in the considered range, all of the total MAEs of SL H-FeX$_2$ (X = Cl, I) experience sensitive change. 

Commonly, the total MAE of SL H-FeX$_2$ (X = Cl, I) shows a rapid decrease after \textit{U$_{\textup{eff}}$} reaches a critical magnitude.
To understand this common feature, the element resolved MAE of SL H-FeX$_2$ (X = Cl, Br, I) under different values of \textit{U$_{\textup{eff}}$} is plotted in  Figs. \ref{fig8}(d)-(f). It can be observed that when \textit{U$_{\textup{eff}}$} is small, the Fe-MAE of SL H-FeX$_2$ (X = Cl, Br, I) remains almost constant. However, after \textit{U$_{\textup{eff}}$} reaches a critical value, the Fe-MAE would suddenly decrease with \textit{U$_{\textup{eff}}$}.   
Similar to the case in SL H-FeBr$_2$, these common features in Fe-MAE of SL H-FeX$_2$ (X = Cl, Br, I) are all originated from the disentanglement between Fe-A$_1$ and E$_1$ orbitals, which are all reflected in the evolution of their band structures with \textit{U$_{\textup{eff}}$} (as shown in Fig. S3 and S5 in the Supplemental Material \cite{SM}). This disentanglement weakens the SOC interaction between occupied Fe- $d_{xy}$ ($d_{x^{2}-y^{2}}$) orbitals at VBM and unoccupied $d_{x^{2}-y^{2}}$ ($d_{xy}$ ) orbitals at CBM, giving rise to a strong decrease of the Fe-MAE. It is worth noting that if the magnetization directions of all SL H-FeX$_2$ (X = Cl, Br, I) compounds are forced to be out-of-plane, the topological phase transitions might commonly happen during the continuous tuning of \textit{U$_{\textup{eff}}$}, giving rise to HVM and QAVH states.

In addition to the common features mentioned above, the MAEs of different members in SL H-FeX$_{2}$ (X = Cl, Br, I) also show significant differences with \textit{U$_{\textup{eff}}$}. First, the MAEs of SL H-FeCl$_{2}$ and H-FeBr$_{2}$ exhibit transitions between positive and negative MAEs with increasing \textit{U$_{\textup{eff}}$}, which corresponds to the reversal of their magnetization behaviors between out-of-plane and in-plane, but the MAE of SL H-FeI$_{2}$ does not behave in this way. Second, when the value of \textit{U$_{\textup{eff}}$} is small, the MAE from X (X=Br,I) elements (denoted as X-MAE) in both SL H-FeBr$_{2}$ and H-FeI$_{2}$ increase significantly with the increase of \textit{U$_{\textup{eff}}$} ; as the \textit{U$_{\textup{eff}}$} value continues to rise, the X-MAEs (X=Br, I) of both systems increase slowly. However, the Cl-MAE of SL H-FeCl$_{2}$ is not sensitive to \textit{U$_{\textup{eff}}$} in the above-mentioned process of increasing \textit{U$_{\textup{eff}}$}, as shown in Figs. \ref{fig8} (d)-(f). Apparently, the halogen atoms of Br and I tend to provide negative values of MAEs. Among them, the amplitude of the negative I-MAE in H-FeI$_{2}$ is larger than that of the positive Fe-MAE, so the total MAE of H-FeI$_{2}$ is always negative. However, in SL H-FeCl$_{2}$, the total MAE of H-FeCl$_{2}$ is almost contributed by Fe-MAE since Cl atoms hardly contribute to MAE. Therefore, the evolution behavior of total MAE in SL H-FeCl$_{2}$ is dominated by Fe-MAE. Based on these calculations, we found that the heavier the halogen atom, the greater the contribution of the halogen atom to the negative MAE.

 Finally, similar with the case of SL H-FeBr$_{2}$, in SL H-FeX$_{2}$ (X = Cl, I) with out-of-plane magnetization, we also observed the chiral edge states connecting the conduction bands and valance bands, as well as an AHC of $-\frac{e^{2}}{h}$ lying in the energy gap (see Fig. S4 and S6 in the Supplemental Material \cite{SM}) in specific intervals of \textit{U$_{\textup{eff}}$}. These characteristics confirm the QAH states with Chern numbers being \textit{C} = -1. We emphasize that these common features of correlation-driven electronic topology stem from the Fe-3d orbitals, which dominates the low-energy states of   SL H-FeX$_{2}$ (X = Cl, Br, I).


\section{\label{sec:level1}CONCLUSION}
We investigated the evolution of MAE as well as electronic structures of SL H-FeBr$_2$ under varying correlation strength, quantified as \textit{U$_{\textup{eff}}$} imposed on Fe ions. It is found that the MAE would decrease after its increase with \textit{U$_{\textup{eff}}$}, and the transition between negative and positive MAE reflects the switching between out-of-plane magnetization and in-plane magnetization. This non-monotonic evolution behaviour of MAE stems from the competition of element-resolved MAE between Fe and Br. The evolution of element-resolved MAE was found to arise from the variation of spin-orbital coupling (SOC) interaction between different orbitals. Further investigation revealed that as  \textit{U$_{\textup{eff}}$} increases, the energy bands of SL H-FeBr$_2$ at K+ and K- invert in turn, giving rise to topological phase transition, and a QAVH state with chiral edge states was predicted. By comparing the MAE evolution behaviours of different members in the SL H-FeX$_2$ (X = Cl, Br, I) family, the underlying mechanism on the universality and specificity of MAE evolution behaviour is provided. Our study has deepened the understanding of correlation-induced electronic structural transition of SL H-FeX$_2$ (X = Cl, Br, I) family , which would open new perspectives of possible spintronics and valleytronics  applications on nanoelectronic devices based on these materials.

\section{\label{sec:level1}ACKNOWLEDGEMENTS}
The author sincerely thanks Wenhui Duan, Yong Xu, Haowei Chen, and Zhiming Xu for helpful discussions.

\appendix

\section{\label{app}The magneto-valley coupling in SL H-FeBr$_2$}

Basically, the coupling of spontaneous valley polarization and magnetization direction stems from the existence of SOC. To illustrate this, we express the SOC Hamiltonian as:
\begin{equation}
 \hat{H}_{\textup{SOC}} = \lambda\hat{L} \cdot \hat{S}     
\end{equation}
Here $\lambda$ represents the strength of SOC, while $\hat{L}$ and $\hat{S}$ are the orbital angular moment and spin angular moment respectively. The SOC can be decoupled as 
\begin{equation}
\hat{H}_{\textup{SOC}} = \hat{H}^{0}_{\textup{SOC}}+\hat{H}^{1}_{SOC}, 
\end{equation}
where $\hat{H}^{0}_{\textup{SOC}}$ represents the interaction between same spin states, and $\hat{H}^{1}_{\textup{SOC}}$ describes the interaction between the states with opposite spin angular moments.
Here $\hat{H}^{0}_{\textup{SOC}}$ and $\hat{H}^{1}_{\textup{SOC}}$ are expressed as\cite{doi:10.1021/acs.inorgchem.9b00687,doi:10.1021/acs.accounts.5b00408}: 
\begin{equation}
   \hat{H}^{0}_{\textup{SOC}}=\lambda \hat{S}_{z'} (\hat{L}_{z} cos\theta + \frac{1}{2} \hat{L}_{+} e^{-i\phi}sin\theta + \frac{1}{2} \hat{L}_{-} e^{i\phi}sin\theta )  
\end{equation}
 
 \begin{equation}
 \begin{aligned}
     \hat{H}^{1}_{\textup{SOC}}&= \frac{\lambda}{2} (\hat{S}_{+'}+\hat{S}_{-'})\\ &\times (-\hat{L}_{z}sin\theta + \frac{1}{2} \hat{L}_{+} e^{-i\phi}cos\theta + \frac{1}{2} \hat{L}_{-} e^{i\phi}cos\theta )     
 \end{aligned}
 \end{equation}
 
 In above expressions, two Cartesian coordinate systems (x, y, z) and (x', y', z') are respectively defined for the orientation of $\hat{L}$ and $\hat{S}$. These two Cartesian coordinate systems are linked by two polar angles, $\theta$ and $\phi$. And
 \begin{equation}
 \hat{L_{\pm}} = \hat{L_{x}} \pm i \hat{L_{y}},
 \end{equation}
 \begin{equation}
 \hat{S_{\pm'}} = \hat{S_{x'}} \pm i \hat{S_{y'}}.
 \end{equation}
 Because both CBM and VBM belong to the same spin channel, only $\hat{H}^{0}_{\textup{SOC}}$ takes effect and $\hat{H}^{1}_{\textup{SOC}}$ can be neglected in the following analyse. 
 
 The orbital components of each valley is plotted in Fig. \ref{fig3}(b). As can be seen, at K+ and K-, the CBM are mainly contributed by Fe-$A_{1}$ orbitals, while the VBM are jointly dominated by Fe-$E_{1}$ orbitals. Moreover, as shown in Fig. \ref{fig3}(a), both CBM and VBM belong to spin-down channel. Therefore, based on the local $C_{3h}$ vector group at K+ and K-, the basis function of CBM and VBM are chosen as:
 \begin{equation}
     |\Psi^{\tau}_{c} \rangle = | d_{z^{2}} \rangle \otimes |\downarrow \rangle 
 \end{equation}
 
  \begin{equation}
     |\Psi^{\tau}_{v} \rangle = \frac{1}{\sqrt{2}}(| d_{x^{2}-y^{2}} \rangle + i\tau d_{xy} \rangle) \otimes |\downarrow \rangle 
 \end{equation}
 where $\tau = \pm 1$ refers to the valley index corresponding to K+/K-. 
 For out of plane magnetization, $\theta$ = 0, we have
 \begin{equation}
 \hat{H}_{\textup{out-of-plane}} \approx \hat{H_{\textup{SOC}}}(\theta = 0) = \lambda\hat{L}_{z} \hat{S}_{z'}. 
 \end{equation}
 As a result, the energy shifting of CBM and VBM caused by SOC are respectively
 \begin{equation}
 E^{\tau}_{c} = \langle \psi^{\tau}_{c} |\hat{H_{\textup{SOC}}}(\theta = 0) | \psi^{\tau}_{c} \rangle 
 \end{equation}
 and 
 \begin{equation}
 E^{\tau}_{v} = \langle \psi^{\tau}_{v} |\hat{H_{\textup{SOC}}}(\theta = 0) | \psi^{\tau}_{v} \rangle, 
 \end{equation}
 the energy difference at either the CBM or the VBM between K+ and K- can be derived as: 
 \begin{equation}
     \Delta E_{c} = E^{K+}_{c}-E^{K-}_{c} = 0
 \end{equation}
 
 \begin{equation}
 \begin{aligned}
     \Delta E_{v}& = E^{K+}_{v}-E^{K-}_{v}\\
     & = i\langle d_{x^{2}-y^{2}} | \hat{H}_{\textup{SOC}} | d_{xy} \rangle - i\langle d_{xy} | \hat{H}_{\textup{SOC}} | d_{x^{2}-y^{2}} \rangle\\
     & = 4\lambda
 \end{aligned}
 \end{equation}
 where we have used \cite{doi:10.1021/acs.inorgchem.9b00687}
 \begin{equation}
 \hat{L}_{z} | d_{x^{2}-y^{2}} \rangle = 2i | d_{xy} \rangle, 
\end{equation}
and
\begin{equation}
 \hat{L}_{z} | d_{xy} \rangle = -2i || d_{x^{2}-y^{2}} \rangle.
\end{equation}
 Thus, the existence of valley polarization has been demonstrated by the existence of nonzero $\Delta E_{v}$ . It can be clearly seen that the energy difference at each valley is solely contributed by SOC valance bands, instead of conduction bands, which is consistent with our calculation results.  
 
 For the case of in-plane magnetization, $\theta$ = $\pi/2$, we have
 \begin{equation}
 \hat{H}_{\textup{in-plane}} \approx \frac{\lambda}{2} (\hat{L}_{+}e^{-i\phi} + \hat{L}_{-}e^{+i\phi})\cdot \hat{S}_{z'}.
 \end{equation}
 In this case, it can be derived that both $\Delta E_{c}$ and $\Delta E_{v}$ are zero, revealing the absence of valley polarization. As a result, mediating by SOC, the valley polarization is coupled with magnetization direction.


\begin{thebibliography}{50}%
\makeatletter
\providecommand \@ifxundefined [1]{%
 \@ifx{#1\undefined}
}%
\providecommand \@ifnum [1]{%
 \ifnum #1\expandafter \@firstoftwo
 \else \expandafter \@secondoftwo
 \fi
}%
\providecommand \@ifx [1]{%
 \ifx #1\expandafter \@firstoftwo
 \else \expandafter \@secondoftwo
 \fi
}%
\providecommand \natexlab [1]{#1}%
\providecommand \enquote  [1]{``#1''}%
\providecommand \bibnamefont  [1]{#1}%
\providecommand \bibfnamefont [1]{#1}%
\providecommand \citenamefont [1]{#1}%
\providecommand \href@noop [0]{\@secondoftwo}%
\providecommand \href [0]{\begingroup \@sanitize@url \@href}%
\providecommand \@href[1]{\@@startlink{#1}\@@href}%
\providecommand \@@href[1]{\endgroup#1\@@endlink}%
\providecommand \@sanitize@url [0]{\catcode `\\12\catcode `\$12\catcode
  `\&12\catcode `\#12\catcode `\^12\catcode `\_12\catcode `\%12\relax}%
\providecommand \@@startlink[1]{}%
\providecommand \@@endlink[0]{}%
\providecommand \url  [0]{\begingroup\@sanitize@url \@url }%
\providecommand \@url [1]{\endgroup\@href {#1}{\urlprefix }}%
\providecommand \urlprefix  [0]{URL }%
\providecommand \Eprint [0]{\href }%
\providecommand \doibase [0]{http://dx.doi.org/}%
\providecommand \selectlanguage [0]{\@gobble}%
\providecommand \bibinfo  [0]{\@secondoftwo}%
\providecommand \bibfield  [0]{\@secondoftwo}%
\providecommand \translation [1]{[#1]}%
\providecommand \BibitemOpen [0]{}%
\providecommand \bibitemStop [0]{}%
\providecommand \bibitemNoStop [0]{.\EOS\space}%
\providecommand \EOS [0]{\spacefactor3000\relax}%
\providecommand \BibitemShut  [1]{\csname bibitem#1\endcsname}%
\let\auto@bib@innerbib\@empty
\bibitem [{\citenamefont {He}\ and\ \citenamefont
  {Xu}(2021)}]{PhysRevB.104.235108}%
  \BibitemOpen
  \bibfield  {author} {\bibinfo {author} {\bibfnamefont {Z.}~\bibnamefont
  {He}}\ and\ \bibinfo {author} {\bibfnamefont {G.}~\bibnamefont {Xu}},\ }\href
  {\doibase 10.1103/PhysRevB.104.235108} {\bibfield  {journal} {\bibinfo
  {journal} {Phys. Rev. B}\ }\textbf {\bibinfo {volume} {104}},\ \bibinfo
  {pages} {235108} (\bibinfo {year} {2021})}\BibitemShut {NoStop}%
\bibitem [{\citenamefont {Wang}\ \emph {et~al.}(2015)\citenamefont {Wang},
  \citenamefont {Wang}, \citenamefont {Fang},\ and\ \citenamefont
  {Dai}}]{PhysRevB.91.125139}%
  \BibitemOpen
  \bibfield  {author} {\bibinfo {author} {\bibfnamefont {Y.}~\bibnamefont
  {Wang}}, \bibinfo {author} {\bibfnamefont {Z.}~\bibnamefont {Wang}}, \bibinfo
  {author} {\bibfnamefont {Z.}~\bibnamefont {Fang}}, \ and\ \bibinfo {author}
  {\bibfnamefont {X.}~\bibnamefont {Dai}},\ }\href {\doibase
  10.1103/PhysRevB.91.125139} {\bibfield  {journal} {\bibinfo  {journal} {Phys.
  Rev. B}\ }\textbf {\bibinfo {volume} {91}},\ \bibinfo {pages} {125139}
  (\bibinfo {year} {2015})}\BibitemShut {NoStop}%
\bibitem [{\citenamefont {Sorella}\ \emph {et~al.}(2018)\citenamefont
  {Sorella}, \citenamefont {Seki}, \citenamefont {Brovko}, \citenamefont
  {Shirakawa}, \citenamefont {Miyakoshi}, \citenamefont {Yunoki},\ and\
  \citenamefont {Tosatti}}]{PhysRevLett.121.066402}%
  \BibitemOpen
  \bibfield  {author} {\bibinfo {author} {\bibfnamefont {S.}~\bibnamefont
  {Sorella}}, \bibinfo {author} {\bibfnamefont {K.}~\bibnamefont {Seki}},
  \bibinfo {author} {\bibfnamefont {O.~O.}\ \bibnamefont {Brovko}}, \bibinfo
  {author} {\bibfnamefont {T.}~\bibnamefont {Shirakawa}}, \bibinfo {author}
  {\bibfnamefont {S.}~\bibnamefont {Miyakoshi}}, \bibinfo {author}
  {\bibfnamefont {S.}~\bibnamefont {Yunoki}}, \ and\ \bibinfo {author}
  {\bibfnamefont {E.}~\bibnamefont {Tosatti}},\ }\href {\doibase
  10.1103/PhysRevLett.121.066402} {\bibfield  {journal} {\bibinfo  {journal}
  {Phys. Rev. Lett.}\ }\textbf {\bibinfo {volume} {121}},\ \bibinfo {pages}
  {066402} (\bibinfo {year} {2018})}\BibitemShut {NoStop}%
\bibitem [{\citenamefont {Chen}\ and\ \citenamefont
  {Lado}(2019)}]{PhysRevLett.122.016803}%
  \BibitemOpen
  \bibfield  {author} {\bibinfo {author} {\bibfnamefont {W.}~\bibnamefont
  {Chen}}\ and\ \bibinfo {author} {\bibfnamefont {J.~L.}\ \bibnamefont
  {Lado}},\ }\href {\doibase 10.1103/PhysRevLett.122.016803} {\bibfield
  {journal} {\bibinfo  {journal} {Phys. Rev. Lett.}\ }\textbf {\bibinfo
  {volume} {122}},\ \bibinfo {pages} {016803} (\bibinfo {year}
  {2019})}\BibitemShut {NoStop}%
\bibitem [{\citenamefont {R\"osner}\ and\ \citenamefont
  {Lado}(2021)}]{PhysRevResearch.3.013265}%
  \BibitemOpen
  \bibfield  {author} {\bibinfo {author} {\bibfnamefont {M.}~\bibnamefont
  {R\"osner}}\ and\ \bibinfo {author} {\bibfnamefont {J.~L.}\ \bibnamefont
  {Lado}},\ }\href {\doibase 10.1103/PhysRevResearch.3.013265} {\bibfield
  {journal} {\bibinfo  {journal} {Phys. Rev. Research}\ }\textbf {\bibinfo
  {volume} {3}},\ \bibinfo {pages} {013265} (\bibinfo {year}
  {2021})}\BibitemShut {NoStop}%
\bibitem [{\citenamefont {Keshavarz}\ \emph {et~al.}(2018)\citenamefont
  {Keshavarz}, \citenamefont {Sch\"ott}, \citenamefont {Millis},\ and\
  \citenamefont {Kvashnin}}]{PhysRevB.97.184404}%
  \BibitemOpen
  \bibfield  {author} {\bibinfo {author} {\bibfnamefont {S.}~\bibnamefont
  {Keshavarz}}, \bibinfo {author} {\bibfnamefont {J.}~\bibnamefont {Sch\"ott}},
  \bibinfo {author} {\bibfnamefont {A.~J.}\ \bibnamefont {Millis}}, \ and\
  \bibinfo {author} {\bibfnamefont {Y.~O.}\ \bibnamefont {Kvashnin}},\ }\href
  {\doibase 10.1103/PhysRevB.97.184404} {\bibfield  {journal} {\bibinfo
  {journal} {Phys. Rev. B}\ }\textbf {\bibinfo {volume} {97}},\ \bibinfo
  {pages} {184404} (\bibinfo {year} {2018})}\BibitemShut {NoStop}%
\bibitem [{\citenamefont {Misumi}\ \emph {et~al.}(2017)\citenamefont {Misumi},
  \citenamefont {Kaneko},\ and\ \citenamefont {Ohta}}]{PhysRevB.95.075124}%
  \BibitemOpen
  \bibfield  {author} {\bibinfo {author} {\bibfnamefont {K.}~\bibnamefont
  {Misumi}}, \bibinfo {author} {\bibfnamefont {T.}~\bibnamefont {Kaneko}}, \
  and\ \bibinfo {author} {\bibfnamefont {Y.}~\bibnamefont {Ohta}},\ }\href
  {\doibase 10.1103/PhysRevB.95.075124} {\bibfield  {journal} {\bibinfo
  {journal} {Phys. Rev. B}\ }\textbf {\bibinfo {volume} {95}},\ \bibinfo
  {pages} {075124} (\bibinfo {year} {2017})}\BibitemShut {NoStop}%
\bibitem [{\citenamefont {Ke}\ and\ \citenamefont
  {Katsnelson}(2021)}]{2021Electron}%
  \BibitemOpen
  \bibfield  {author} {\bibinfo {author} {\bibfnamefont {L.}~\bibnamefont
  {Ke}}\ and\ \bibinfo {author} {\bibfnamefont {M.~I.}\ \bibnamefont
  {Katsnelson}},\ }\href {https://www.nature.com/articles/s41524-020-00469-2}
  {\bibfield  {journal} {\bibinfo  {journal} {npj Computational Materials}\
  }\textbf {\bibinfo {volume} {7}},\ \bibinfo {pages} {1} (\bibinfo {year}
  {2021})}\BibitemShut {NoStop}%
\bibitem [{\citenamefont {Gray}\ \emph {et~al.}(2016)\citenamefont {Gray},
  \citenamefont {Jeong}, \citenamefont {Aetukuri}, \citenamefont {Granitzka},
  \citenamefont {Chen}, \citenamefont {Kukreja}, \citenamefont {Higley},
  \citenamefont {Chase}, \citenamefont {Reid}, \citenamefont {Ohldag},
  \citenamefont {Marcus}, \citenamefont {Scholl}, \citenamefont {Young},
  \citenamefont {Doran}, \citenamefont {Jenkins}, \citenamefont {Shafer},
  \citenamefont {Arenholz}, \citenamefont {Samant}, \citenamefont {Parkin},\
  and\ \citenamefont {D\"urr}}]{PhysRevLett.116.116403}%
  \BibitemOpen
  \bibfield  {author} {\bibinfo {author} {\bibfnamefont {A.~X.}\ \bibnamefont
  {Gray}}, \bibinfo {author} {\bibfnamefont {J.}~\bibnamefont {Jeong}},
  \bibinfo {author} {\bibfnamefont {N.~P.}\ \bibnamefont {Aetukuri}}, \bibinfo
  {author} {\bibfnamefont {P.}~\bibnamefont {Granitzka}}, \bibinfo {author}
  {\bibfnamefont {Z.}~\bibnamefont {Chen}}, \bibinfo {author} {\bibfnamefont
  {R.}~\bibnamefont {Kukreja}}, \bibinfo {author} {\bibfnamefont
  {D.}~\bibnamefont {Higley}}, \bibinfo {author} {\bibfnamefont
  {T.}~\bibnamefont {Chase}}, \bibinfo {author} {\bibfnamefont {A.~H.}\
  \bibnamefont {Reid}}, \bibinfo {author} {\bibfnamefont {H.}~\bibnamefont
  {Ohldag}}, \bibinfo {author} {\bibfnamefont {M.~A.}\ \bibnamefont {Marcus}},
  \bibinfo {author} {\bibfnamefont {A.}~\bibnamefont {Scholl}}, \bibinfo
  {author} {\bibfnamefont {A.~T.}\ \bibnamefont {Young}}, \bibinfo {author}
  {\bibfnamefont {A.}~\bibnamefont {Doran}}, \bibinfo {author} {\bibfnamefont
  {C.~A.}\ \bibnamefont {Jenkins}}, \bibinfo {author} {\bibfnamefont
  {P.}~\bibnamefont {Shafer}}, \bibinfo {author} {\bibfnamefont
  {E.}~\bibnamefont {Arenholz}}, \bibinfo {author} {\bibfnamefont {M.~G.}\
  \bibnamefont {Samant}}, \bibinfo {author} {\bibfnamefont {S.~S.~P.}\
  \bibnamefont {Parkin}}, \ and\ \bibinfo {author} {\bibfnamefont {H.~A.}\
  \bibnamefont {D\"urr}},\ }\href {\doibase 10.1103/PhysRevLett.116.116403}
  {\bibfield  {journal} {\bibinfo  {journal} {Phys. Rev. Lett.}\ }\textbf
  {\bibinfo {volume} {116}},\ \bibinfo {pages} {116403} (\bibinfo {year}
  {2016})}\BibitemShut {NoStop}%
\bibitem [{\citenamefont {De~Franco}\ \emph {et~al.}(2018)\citenamefont
  {De~Franco}, \citenamefont {Tocchio},\ and\ \citenamefont
  {Becca}}]{PhysRevB.98.075117}%
  \BibitemOpen
  \bibfield  {author} {\bibinfo {author} {\bibfnamefont {C.}~\bibnamefont
  {De~Franco}}, \bibinfo {author} {\bibfnamefont {L.~F.}\ \bibnamefont
  {Tocchio}}, \ and\ \bibinfo {author} {\bibfnamefont {F.}~\bibnamefont
  {Becca}},\ }\href {\doibase 10.1103/PhysRevB.98.075117} {\bibfield  {journal}
  {\bibinfo  {journal} {Phys. Rev. B}\ }\textbf {\bibinfo {volume} {98}},\
  \bibinfo {pages} {075117} (\bibinfo {year} {2018})}\BibitemShut {NoStop}%
\bibitem [{\citenamefont {Yao}\ \emph {et~al.}(2021)\citenamefont {Yao},
  \citenamefont {Li},\ and\ \citenamefont {Liu}}]{PhysRevB.104.035108}%
  \BibitemOpen
  \bibfield  {author} {\bibinfo {author} {\bibfnamefont {Q.}~\bibnamefont
  {Yao}}, \bibinfo {author} {\bibfnamefont {J.}~\bibnamefont {Li}}, \ and\
  \bibinfo {author} {\bibfnamefont {Q.}~\bibnamefont {Liu}},\ }\href {\doibase
  10.1103/PhysRevB.104.035108} {\bibfield  {journal} {\bibinfo  {journal}
  {Phys. Rev. B}\ }\textbf {\bibinfo {volume} {104}},\ \bibinfo {pages}
  {035108} (\bibinfo {year} {2021})}\BibitemShut {NoStop}%
\bibitem [{\citenamefont {Raja}\ \emph {et~al.}(2017)\citenamefont {Raja},
  \citenamefont {Chaves}, \citenamefont {Yu}, \citenamefont {Arefe},
  \citenamefont {Hill}, \citenamefont {Rigosi}, \citenamefont {Berkelbach},
  \citenamefont {Nagler}, \citenamefont {Schüller},\ and\ \citenamefont
  {Korn}}]{2017Coulomb}%
  \BibitemOpen
  \bibfield  {author} {\bibinfo {author} {\bibfnamefont {A.}~\bibnamefont
  {Raja}}, \bibinfo {author} {\bibfnamefont {A.}~\bibnamefont {Chaves}},
  \bibinfo {author} {\bibfnamefont {J.}~\bibnamefont {Yu}}, \bibinfo {author}
  {\bibfnamefont {G.}~\bibnamefont {Arefe}}, \bibinfo {author} {\bibfnamefont
  {H.~M.}\ \bibnamefont {Hill}}, \bibinfo {author} {\bibfnamefont
  {A.}~\bibnamefont {Rigosi}}, \bibinfo {author} {\bibfnamefont {T.~C.}\
  \bibnamefont {Berkelbach}}, \bibinfo {author} {\bibfnamefont
  {P.}~\bibnamefont {Nagler}}, \bibinfo {author} {\bibfnamefont
  {C.}~\bibnamefont {Schüller}}, \ and\ \bibinfo {author} {\bibfnamefont
  {T.}~\bibnamefont {Korn}},\ }\href {https://doi.org/10.1038/ncomms15251}
  {\bibfield  {journal} {\bibinfo  {journal} {Nature Communications}\ }\textbf
  {\bibinfo {volume} {8}},\ \bibinfo {pages} {15251} (\bibinfo {year}
  {2017})}\BibitemShut {NoStop}%
\bibitem [{\citenamefont {Utama}\ \emph {et~al.}(2019)\citenamefont {Utama},
  \citenamefont {Kleemann}, \citenamefont {Zhao}, \citenamefont {Ong},
  \citenamefont {da~Jornada}, \citenamefont {Qiu}, \citenamefont {Cai},
  \citenamefont {Li}, \citenamefont {Kou}, \citenamefont {Zhao} \emph
  {et~al.}}]{2019A}%
  \BibitemOpen
  \bibfield  {author} {\bibinfo {author} {\bibfnamefont {M.}~\bibnamefont
  {Utama}}, \bibinfo {author} {\bibfnamefont {H.}~\bibnamefont {Kleemann}},
  \bibinfo {author} {\bibfnamefont {W.}~\bibnamefont {Zhao}}, \bibinfo {author}
  {\bibfnamefont {C.~S.}\ \bibnamefont {Ong}}, \bibinfo {author} {\bibfnamefont
  {F.~H.}\ \bibnamefont {da~Jornada}}, \bibinfo {author} {\bibfnamefont
  {D.~Y.}\ \bibnamefont {Qiu}}, \bibinfo {author} {\bibfnamefont
  {H.}~\bibnamefont {Cai}}, \bibinfo {author} {\bibfnamefont {H.}~\bibnamefont
  {Li}}, \bibinfo {author} {\bibfnamefont {R.}~\bibnamefont {Kou}}, \bibinfo
  {author} {\bibfnamefont {S.}~\bibnamefont {Zhao}},  \emph {et~al.},\ }\href
  {https://www.nature.com/articles/s41928-019-0207-4} {\bibfield  {journal}
  {\bibinfo  {journal} {Nature Electronics}\ }\textbf {\bibinfo {volume} {2}},\
  \bibinfo {pages} {60} (\bibinfo {year} {2019})}\BibitemShut {NoStop}%
\bibitem [{\citenamefont {Waldecker}\ \emph {et~al.}(2019)\citenamefont
  {Waldecker}, \citenamefont {Raja}, \citenamefont {R\"osner}, \citenamefont
  {Steinke}, \citenamefont {Bostwick}, \citenamefont {Koch}, \citenamefont
  {Jozwiak}, \citenamefont {Taniguchi}, \citenamefont {Watanabe}, \citenamefont
  {Rotenberg}, \citenamefont {Wehling},\ and\ \citenamefont
  {Heinz}}]{PhysRevLett.123.206403}%
  \BibitemOpen
  \bibfield  {author} {\bibinfo {author} {\bibfnamefont {L.}~\bibnamefont
  {Waldecker}}, \bibinfo {author} {\bibfnamefont {A.}~\bibnamefont {Raja}},
  \bibinfo {author} {\bibfnamefont {M.}~\bibnamefont {R\"osner}}, \bibinfo
  {author} {\bibfnamefont {C.}~\bibnamefont {Steinke}}, \bibinfo {author}
  {\bibfnamefont {A.}~\bibnamefont {Bostwick}}, \bibinfo {author}
  {\bibfnamefont {R.~J.}\ \bibnamefont {Koch}}, \bibinfo {author}
  {\bibfnamefont {C.}~\bibnamefont {Jozwiak}}, \bibinfo {author} {\bibfnamefont
  {T.}~\bibnamefont {Taniguchi}}, \bibinfo {author} {\bibfnamefont
  {K.}~\bibnamefont {Watanabe}}, \bibinfo {author} {\bibfnamefont
  {E.}~\bibnamefont {Rotenberg}}, \bibinfo {author} {\bibfnamefont {T.~O.}\
  \bibnamefont {Wehling}}, \ and\ \bibinfo {author} {\bibfnamefont {T.~F.}\
  \bibnamefont {Heinz}},\ }\href {\doibase 10.1103/PhysRevLett.123.206403}
  {\bibfield  {journal} {\bibinfo  {journal} {Phys. Rev. Lett.}\ }\textbf
  {\bibinfo {volume} {123}},\ \bibinfo {pages} {206403} (\bibinfo {year}
  {2019})}\BibitemShut {NoStop}%
\bibitem [{\citenamefont {Meng}\ \emph {et~al.}(2021)\citenamefont {Meng},
  \citenamefont {Zhou}, \citenamefont {Xu}, \citenamefont {Yang},\ and\
  \citenamefont {Gong}}]{2021Anomalous}%
  \BibitemOpen
  \bibfield  {author} {\bibinfo {author} {\bibfnamefont {L.}~\bibnamefont
  {Meng}}, \bibinfo {author} {\bibfnamefont {Z.}~\bibnamefont {Zhou}}, \bibinfo
  {author} {\bibfnamefont {M.}~\bibnamefont {Xu}}, \bibinfo {author}
  {\bibfnamefont {S.}~\bibnamefont {Yang}}, \ and\ \bibinfo {author}
  {\bibfnamefont {Y.}~\bibnamefont {Gong}},\ }\href
  {https://doi.org/10.1038/s41467-021-21072-z} {\bibfield  {journal} {\bibinfo
  {journal} {Nature Communications}\ }\textbf {\bibinfo {volume} {12}},\
  \bibinfo {pages} {809} (\bibinfo {year} {2021})}\BibitemShut {NoStop}%
\bibitem [{\citenamefont {Karbalaee~Aghaee}\ \emph {et~al.}(2022)\citenamefont
  {Karbalaee~Aghaee}, \citenamefont {Belbasi},\ and\ \citenamefont
  {Hadipour}}]{PhysRevB.105.115115}%
  \BibitemOpen
  \bibfield  {author} {\bibinfo {author} {\bibfnamefont {A.}~\bibnamefont
  {Karbalaee~Aghaee}}, \bibinfo {author} {\bibfnamefont {S.}~\bibnamefont
  {Belbasi}}, \ and\ \bibinfo {author} {\bibfnamefont {H.}~\bibnamefont
  {Hadipour}},\ }\href {\doibase 10.1103/PhysRevB.105.115115} {\bibfield
  {journal} {\bibinfo  {journal} {Phys. Rev. B}\ }\textbf {\bibinfo {volume}
  {105}},\ \bibinfo {pages} {115115} (\bibinfo {year} {2022})}\BibitemShut
  {NoStop}%
\bibitem [{\citenamefont {Li}\ \emph {et~al.}(2021)\citenamefont {Li},
  \citenamefont {Wang}, \citenamefont {Zhang}, \citenamefont {Guo},\ and\
  \citenamefont {Yang}}]{PhysRevB.104.085149}%
  \BibitemOpen
  \bibfield  {author} {\bibinfo {author} {\bibfnamefont {S.}~\bibnamefont
  {Li}}, \bibinfo {author} {\bibfnamefont {Q.}~\bibnamefont {Wang}}, \bibinfo
  {author} {\bibfnamefont {C.}~\bibnamefont {Zhang}}, \bibinfo {author}
  {\bibfnamefont {P.}~\bibnamefont {Guo}}, \ and\ \bibinfo {author}
  {\bibfnamefont {S.~A.}\ \bibnamefont {Yang}},\ }\href {\doibase
  10.1103/PhysRevB.104.085149} {\bibfield  {journal} {\bibinfo  {journal}
  {Phys. Rev. B}\ }\textbf {\bibinfo {volume} {104}},\ \bibinfo {pages}
  {085149} (\bibinfo {year} {2021})}\BibitemShut {NoStop}%
\bibitem [{\citenamefont {Kong}\ \emph {et~al.}(2020)\citenamefont {Kong},
  \citenamefont {Li}, \citenamefont {Liang}, \citenamefont {Peeters},\ and\
  \citenamefont {Liu}}]{2020The}%
  \BibitemOpen
  \bibfield  {author} {\bibinfo {author} {\bibfnamefont {X.}~\bibnamefont
  {Kong}}, \bibinfo {author} {\bibfnamefont {L.}~\bibnamefont {Li}}, \bibinfo
  {author} {\bibfnamefont {L.}~\bibnamefont {Liang}}, \bibinfo {author}
  {\bibfnamefont {F.~M.}\ \bibnamefont {Peeters}}, \ and\ \bibinfo {author}
  {\bibfnamefont {X.~J.}\ \bibnamefont {Liu}},\ }\href
  {https://aip.scitation.org/doi/10.1063/5.0006446} {\bibfield  {journal}
  {\bibinfo  {journal} {Applied Physics Letters}\ }\textbf {\bibinfo {volume}
  {116}},\ \bibinfo {pages} {192404} (\bibinfo {year} {2020})}\BibitemShut
  {NoStop}%
\bibitem [{\citenamefont {Hu}\ \emph {et~al.}(2020)\citenamefont {Hu},
  \citenamefont {Tong}, \citenamefont {Shen}, \citenamefont {Wan},\ and\
  \citenamefont {Duan}}]{hu2020concepts}%
  \BibitemOpen
  \bibfield  {author} {\bibinfo {author} {\bibfnamefont {H.}~\bibnamefont
  {Hu}}, \bibinfo {author} {\bibfnamefont {W.-Y.}\ \bibnamefont {Tong}},
  \bibinfo {author} {\bibfnamefont {Y.-H.}\ \bibnamefont {Shen}}, \bibinfo
  {author} {\bibfnamefont {X.}~\bibnamefont {Wan}}, \ and\ \bibinfo {author}
  {\bibfnamefont {C.-G.}\ \bibnamefont {Duan}},\ }\href
  {https://doi.org/10.1038/s41524-020-00397-1} {\bibfield  {journal} {\bibinfo
  {journal} {npj Computational Materials}\ }\textbf {\bibinfo {volume} {6}},\
  \bibinfo {pages} {129} (\bibinfo {year} {2020})}\BibitemShut {NoStop}%
\bibitem [{\citenamefont {Zhao}\ \emph {et~al.}(2022)\citenamefont {Zhao},
  \citenamefont {Dai}, \citenamefont {Wang}, \citenamefont {Huang},\ and\
  \citenamefont {Ma}}]{ZHAO202256}%
  \BibitemOpen
  \bibfield  {author} {\bibinfo {author} {\bibfnamefont {P.}~\bibnamefont
  {Zhao}}, \bibinfo {author} {\bibfnamefont {Y.}~\bibnamefont {Dai}}, \bibinfo
  {author} {\bibfnamefont {H.}~\bibnamefont {Wang}}, \bibinfo {author}
  {\bibfnamefont {B.}~\bibnamefont {Huang}}, \ and\ \bibinfo {author}
  {\bibfnamefont {Y.}~\bibnamefont {Ma}},\ }\href {\doibase
  https://doi.org/10.1016/j.chphma.2021.09.006} {\bibfield  {journal} {\bibinfo
   {journal} {ChemPhysMater}\ }\textbf {\bibinfo {volume} {1}},\ \bibinfo
  {pages} {56} (\bibinfo {year} {2022})}\BibitemShut {NoStop}%
\bibitem [{\citenamefont {Tong}\ \emph {et~al.}(2016)\citenamefont {Tong},
  \citenamefont {Gong}, \citenamefont {Wan},\ and\ \citenamefont
  {Duan}}]{2016Concepts}%
  \BibitemOpen
  \bibfield  {author} {\bibinfo {author} {\bibfnamefont {W.~Y.}\ \bibnamefont
  {Tong}}, \bibinfo {author} {\bibfnamefont {S.~J.}\ \bibnamefont {Gong}},
  \bibinfo {author} {\bibfnamefont {X.}~\bibnamefont {Wan}}, \ and\ \bibinfo
  {author} {\bibfnamefont {C.~G.}\ \bibnamefont {Duan}},\ }\href
  {https://doi.org/10.1038/ncomms13612} {\bibfield  {journal} {\bibinfo
  {journal} {Nature Communications}\ }\textbf {\bibinfo {volume} {7}},\
  \bibinfo {pages} {13612} (\bibinfo {year} {2016})}\BibitemShut {NoStop}%
\bibitem [{\citenamefont {Guo}\ \emph {et~al.}(2022)\citenamefont {Guo},
  \citenamefont {Zhu}, \citenamefont {Yin},\ and\ \citenamefont
  {Liu}}]{PhysRevB.105.104416}%
  \BibitemOpen
  \bibfield  {author} {\bibinfo {author} {\bibfnamefont {S.-D.}\ \bibnamefont
  {Guo}}, \bibinfo {author} {\bibfnamefont {J.-X.}\ \bibnamefont {Zhu}},
  \bibinfo {author} {\bibfnamefont {M.-Y.}\ \bibnamefont {Yin}}, \ and\
  \bibinfo {author} {\bibfnamefont {B.-G.}\ \bibnamefont {Liu}},\ }\href
  {\doibase 10.1103/PhysRevB.105.104416} {\bibfield  {journal} {\bibinfo
  {journal} {Phys. Rev. B}\ }\textbf {\bibinfo {volume} {105}},\ \bibinfo
  {pages} {104416} (\bibinfo {year} {2022})}\BibitemShut {NoStop}%
\bibitem [{\citenamefont {Peng}\ \emph {et~al.}(2020)\citenamefont {Peng},
  \citenamefont {Ma}, \citenamefont {Xu}, \citenamefont {He}, \citenamefont
  {Huang},\ and\ \citenamefont {Dai}}]{PhysRevB.102.035412}%
  \BibitemOpen
  \bibfield  {author} {\bibinfo {author} {\bibfnamefont {R.}~\bibnamefont
  {Peng}}, \bibinfo {author} {\bibfnamefont {Y.}~\bibnamefont {Ma}}, \bibinfo
  {author} {\bibfnamefont {X.}~\bibnamefont {Xu}}, \bibinfo {author}
  {\bibfnamefont {Z.}~\bibnamefont {He}}, \bibinfo {author} {\bibfnamefont
  {B.}~\bibnamefont {Huang}}, \ and\ \bibinfo {author} {\bibfnamefont
  {Y.}~\bibnamefont {Dai}},\ }\href {\doibase 10.1103/PhysRevB.102.035412}
  {\bibfield  {journal} {\bibinfo  {journal} {Phys. Rev. B}\ }\textbf {\bibinfo
  {volume} {102}},\ \bibinfo {pages} {035412} (\bibinfo {year}
  {2020})}\BibitemShut {NoStop}%
\bibitem [{\citenamefont {Kresse}\ and\ \citenamefont
  {Furthm\"uller}(1996)}]{PhysRevB.54.11169}%
  \BibitemOpen
  \bibfield  {author} {\bibinfo {author} {\bibfnamefont {G.}~\bibnamefont
  {Kresse}}\ and\ \bibinfo {author} {\bibfnamefont {J.}~\bibnamefont
  {Furthm\"uller}},\ }\href {\doibase 10.1103/PhysRevB.54.11169} {\bibfield
  {journal} {\bibinfo  {journal} {Phys. Rev. B}\ }\textbf {\bibinfo {volume}
  {54}},\ \bibinfo {pages} {11169} (\bibinfo {year} {1996})}\BibitemShut
  {NoStop}%
\bibitem [{\citenamefont {Perdew}\ \emph {et~al.}(1996)\citenamefont {Perdew},
  \citenamefont {Burke},\ and\ \citenamefont
  {Ernzerhof}}]{PhysRevLett.77.3865}%
  \BibitemOpen
  \bibfield  {author} {\bibinfo {author} {\bibfnamefont {J.~P.}\ \bibnamefont
  {Perdew}}, \bibinfo {author} {\bibfnamefont {K.}~\bibnamefont {Burke}}, \
  and\ \bibinfo {author} {\bibfnamefont {M.}~\bibnamefont {Ernzerhof}},\ }\href
  {\doibase 10.1103/PhysRevLett.77.3865} {\bibfield  {journal} {\bibinfo
  {journal} {Phys. Rev. Lett.}\ }\textbf {\bibinfo {volume} {77}},\ \bibinfo
  {pages} {3865} (\bibinfo {year} {1996})}\BibitemShut {NoStop}%
\bibitem [{\citenamefont {Dudarev}\ \emph {et~al.}(1998)\citenamefont
  {Dudarev}, \citenamefont {Botton}, \citenamefont {Savrasov}, \citenamefont
  {Humphreys},\ and\ \citenamefont {Sutton}}]{PhysRevB.57.1505}%
  \BibitemOpen
  \bibfield  {author} {\bibinfo {author} {\bibfnamefont {S.~L.}\ \bibnamefont
  {Dudarev}}, \bibinfo {author} {\bibfnamefont {G.~A.}\ \bibnamefont {Botton}},
  \bibinfo {author} {\bibfnamefont {S.~Y.}\ \bibnamefont {Savrasov}}, \bibinfo
  {author} {\bibfnamefont {C.~J.}\ \bibnamefont {Humphreys}}, \ and\ \bibinfo
  {author} {\bibfnamefont {A.~P.}\ \bibnamefont {Sutton}},\ }\href {\doibase
  10.1103/PhysRevB.57.1505} {\bibfield  {journal} {\bibinfo  {journal} {Phys.
  Rev. B}\ }\textbf {\bibinfo {volume} {57}},\ \bibinfo {pages} {1505}
  (\bibinfo {year} {1998})}\BibitemShut {NoStop}%
\bibitem [{\citenamefont {Grimme}\ \emph {et~al.}(2010)\citenamefont {Grimme},
  \citenamefont {Antony}, \citenamefont {Ehrlich},\ and\ \citenamefont
  {Krieg}}]{2010A}%
  \BibitemOpen
  \bibfield  {author} {\bibinfo {author} {\bibfnamefont {S.}~\bibnamefont
  {Grimme}}, \bibinfo {author} {\bibfnamefont {J.}~\bibnamefont {Antony}},
  \bibinfo {author} {\bibfnamefont {S.}~\bibnamefont {Ehrlich}}, \ and\
  \bibinfo {author} {\bibfnamefont {H.}~\bibnamefont {Krieg}},\ }\href
  {https://aip.scitation.org/doi/10.1063/1.3382344} {\bibfield  {journal}
  {\bibinfo  {journal} {Journal of Chemical Physics}\ }\textbf {\bibinfo
  {volume} {132}},\ \bibinfo {pages} {154104} (\bibinfo {year}
  {2010})}\BibitemShut {NoStop}%
\bibitem [{\citenamefont {Mostofi}\ \emph {et~al.}(2014)\citenamefont
  {Mostofi}, \citenamefont {Yates}, \citenamefont {Pizzi}, \citenamefont {Lee},
  \citenamefont {Souza}, \citenamefont {Vanderbilt},\ and\ \citenamefont
  {Marzari}}]{MOSTOFI20142309}%
  \BibitemOpen
  \bibfield  {author} {\bibinfo {author} {\bibfnamefont {A.~A.}\ \bibnamefont
  {Mostofi}}, \bibinfo {author} {\bibfnamefont {J.~R.}\ \bibnamefont {Yates}},
  \bibinfo {author} {\bibfnamefont {G.}~\bibnamefont {Pizzi}}, \bibinfo
  {author} {\bibfnamefont {Y.-S.}\ \bibnamefont {Lee}}, \bibinfo {author}
  {\bibfnamefont {I.}~\bibnamefont {Souza}}, \bibinfo {author} {\bibfnamefont
  {D.}~\bibnamefont {Vanderbilt}}, \ and\ \bibinfo {author} {\bibfnamefont
  {N.}~\bibnamefont {Marzari}},\ }\href {\doibase
  https://doi.org/10.1016/j.cpc.2014.05.003} {\bibfield  {journal} {\bibinfo
  {journal} {Computer Physics Communications}\ }\textbf {\bibinfo {volume}
  {185}},\ \bibinfo {pages} {2309} (\bibinfo {year} {2014})}\BibitemShut
  {NoStop}%
\bibitem [{\citenamefont {Wu}\ \emph {et~al.}(2018)\citenamefont {Wu},
  \citenamefont {Zhang}, \citenamefont {Song}, \citenamefont {Troyer},\ and\
  \citenamefont {Soluyanov}}]{2017WannierTools}%
  \BibitemOpen
  \bibfield  {author} {\bibinfo {author} {\bibfnamefont {Q.~S.}\ \bibnamefont
  {Wu}}, \bibinfo {author} {\bibfnamefont {S.~N.}\ \bibnamefont {Zhang}},
  \bibinfo {author} {\bibfnamefont {H.~F.}\ \bibnamefont {Song}}, \bibinfo
  {author} {\bibfnamefont {M.}~\bibnamefont {Troyer}}, \ and\ \bibinfo {author}
  {\bibfnamefont {A.~A.}\ \bibnamefont {Soluyanov}},\ }\href
  {https://www.sciencedirect.com/science/article/pii/S0010465517303442}
  {\bibfield  {journal} {\bibinfo  {journal} {Computer Physics Communications}\
  }\textbf {\bibinfo {volume} {224}},\ \bibinfo {pages} {405} (\bibinfo {year}
  {2018})}\BibitemShut {NoStop}%
\bibitem [{\citenamefont {ul~ain}\ \emph {et~al.}(2020)\citenamefont {ul~ain},
  \citenamefont {Odkhuu}, \citenamefont {Rhim},\ and\ \citenamefont
  {Hong}}]{PhysRevB.101.214436}%
  \BibitemOpen
  \bibfield  {author} {\bibinfo {author} {\bibfnamefont {Qurat-ul-ain}~}, \bibinfo {author} {\bibfnamefont {D.}~\bibnamefont {Odkhuu}},
  \bibinfo {author} {\bibfnamefont {S.~H.}\ \bibnamefont {Rhim}}, \ and\
  \bibinfo {author} {\bibfnamefont {S.~C.}\ \bibnamefont {Hong}},\ }\href
  {\doibase 10.1103/PhysRevB.101.214436} {\bibfield  {journal} {\bibinfo
  {journal} {Phys. Rev. B}\ }\textbf {\bibinfo {volume} {101}},\ \bibinfo
  {pages} {214436} (\bibinfo {year} {2020})}\BibitemShut {NoStop}%
\bibitem [{\citenamefont {Antropov}\ \emph {et~al.}(2014)\citenamefont
  {Antropov}, \citenamefont {Ke},\ and\ \citenamefont
  {Åberg}}]{ANTROPOV201435}%
  \BibitemOpen
  \bibfield  {author} {\bibinfo {author} {\bibfnamefont {V.}~\bibnamefont
  {Antropov}}, \bibinfo {author} {\bibfnamefont {L.}~\bibnamefont {Ke}}, \ and\
  \bibinfo {author} {\bibfnamefont {D.}~\bibnamefont {Åberg}},\ }\href
  {\doibase https://doi.org/10.1016/j.ssc.2014.06.003} {\bibfield  {journal}
  {\bibinfo  {journal} {Solid State Communications}\ }\textbf {\bibinfo
  {volume} {194}},\ \bibinfo {pages} {35} (\bibinfo {year} {2014})}\BibitemShut
  {NoStop}%
\bibitem [{\citenamefont {Zhang}\ \emph {et~al.}(2017)\citenamefont {Zhang},
  \citenamefont {Lukashev}, \citenamefont {Jaswal},\ and\ \citenamefont
  {Tsymbal}}]{PhysRevB.96.014435}%
  \BibitemOpen
  \bibfield  {author} {\bibinfo {author} {\bibfnamefont {J.}~\bibnamefont
  {Zhang}}, \bibinfo {author} {\bibfnamefont {P.~V.}\ \bibnamefont {Lukashev}},
  \bibinfo {author} {\bibfnamefont {S.~S.}\ \bibnamefont {Jaswal}}, \ and\
  \bibinfo {author} {\bibfnamefont {E.~Y.}\ \bibnamefont {Tsymbal}},\ }\href
  {\doibase 10.1103/PhysRevB.96.014435} {\bibfield  {journal} {\bibinfo
  {journal} {Phys. Rev. B}\ }\textbf {\bibinfo {volume} {96}},\ \bibinfo
  {pages} {014435} (\bibinfo {year} {2017})}\BibitemShut {NoStop}%
\bibitem [{\citenamefont {Skomski}\ \emph {et~al.}(2011)\citenamefont
  {Skomski}, \citenamefont {Kashyap},\ and\ \citenamefont {Enders}}]{2011Is}%
  \BibitemOpen
  \bibfield  {author} {\bibinfo {author} {\bibfnamefont {R.}~\bibnamefont
  {Skomski}}, \bibinfo {author} {\bibfnamefont {A.}~\bibnamefont {Kashyap}}, \
  and\ \bibinfo {author} {\bibfnamefont {A.}~\bibnamefont {Enders}},\ }\href
  {https://aip.scitation.org/doi/10.1063/1.3562445} {\bibfield  {journal}
  {\bibinfo  {journal} {Journal of Applied Physics}\ }\textbf {\bibinfo
  {volume} {109}},\ \bibinfo {pages} {07E143} (\bibinfo {year}
  {2011})}\BibitemShut {NoStop}%
\bibitem [{\citenamefont {Goodenough}(1955)}]{PhysRev.100.564}%
  \BibitemOpen
  \bibfield  {author} {\bibinfo {author} {\bibfnamefont {J.~B.}\ \bibnamefont
  {Goodenough}},\ }\href {\doibase 10.1103/PhysRev.100.564} {\bibfield
  {journal} {\bibinfo  {journal} {Phys. Rev.}\ }\textbf {\bibinfo {volume}
  {100}},\ \bibinfo {pages} {564} (\bibinfo {year} {1955})}\BibitemShut
  {NoStop}%
\bibitem [{\citenamefont {Kanamori}(1959)}]{KANAMORI195987}%
  \BibitemOpen
  \bibfield  {author} {\bibinfo {author} {\bibfnamefont {J.}~\bibnamefont
  {Kanamori}},\ }\href {\doibase https://doi.org/10.1016/0022-3697(59)90061-7}
  {\bibfield  {journal} {\bibinfo  {journal} {Journal of Physics and Chemistry
  of Solids}\ }\textbf {\bibinfo {volume} {10}},\ \bibinfo {pages} {87}
  (\bibinfo {year} {1959})}\BibitemShut {NoStop}%
\bibitem [{\citenamefont {Xu}\ \emph {et~al.}(2022)\citenamefont {Xu},
  \citenamefont {Ye}, \citenamefont {Li}, \citenamefont {Duan},\ and\
  \citenamefont {Xu}}]{PhysRevB.105.085129}%
  \BibitemOpen
  \bibfield  {author} {\bibinfo {author} {\bibfnamefont {Z.}~\bibnamefont
  {Xu}}, \bibinfo {author} {\bibfnamefont {M.}~\bibnamefont {Ye}}, \bibinfo
  {author} {\bibfnamefont {J.}~\bibnamefont {Li}}, \bibinfo {author}
  {\bibfnamefont {W.}~\bibnamefont {Duan}}, \ and\ \bibinfo {author}
  {\bibfnamefont {Y.}~\bibnamefont {Xu}},\ }\href {\doibase
  10.1103/PhysRevB.105.085129} {\bibfield  {journal} {\bibinfo  {journal}
  {Phys. Rev. B}\ }\textbf {\bibinfo {volume} {105}},\ \bibinfo {pages}
  {085129} (\bibinfo {year} {2022})}\BibitemShut {NoStop}%
\bibitem [{\citenamefont {Wang}\ \emph {et~al.}(1993)\citenamefont {Wang},
  \citenamefont {Wu},\ and\ \citenamefont {Freeman}}]{PhysRevB.47.14932}%
  \BibitemOpen
  \bibfield  {author} {\bibinfo {author} {\bibfnamefont {D.-s.}\ \bibnamefont
  {Wang}}, \bibinfo {author} {\bibfnamefont {R.}~\bibnamefont {Wu}}, \ and\
  \bibinfo {author} {\bibfnamefont {A.~J.}\ \bibnamefont {Freeman}},\ }\href
  {\doibase 10.1103/PhysRevB.47.14932} {\bibfield  {journal} {\bibinfo
  {journal} {Phys. Rev. B}\ }\textbf {\bibinfo {volume} {47}},\ \bibinfo
  {pages} {14932} (\bibinfo {year} {1993})}\BibitemShut {NoStop}%
\bibitem [{\citenamefont {Chen}\ \emph {et~al.}(2021)\citenamefont {Chen},
  \citenamefont {Tang},\ and\ \citenamefont {Li}}]{PhysRevB.103.195402}%
  \BibitemOpen
  \bibfield  {author} {\bibinfo {author} {\bibfnamefont {H.}~\bibnamefont
  {Chen}}, \bibinfo {author} {\bibfnamefont {P.}~\bibnamefont {Tang}}, \ and\
  \bibinfo {author} {\bibfnamefont {J.}~\bibnamefont {Li}},\ }\href {\doibase
  10.1103/PhysRevB.103.195402} {\bibfield  {journal} {\bibinfo  {journal}
  {Phys. Rev. B}\ }\textbf {\bibinfo {volume} {103}},\ \bibinfo {pages}
  {195402} (\bibinfo {year} {2021})}\BibitemShut {NoStop}%
\bibitem [{SM()}]{SM}%
  \BibitemOpen
  \href@noop {} {\enquote {\bibinfo {title} {See {$\mathrm{Supplemental}$}
  {$\mathrm{Material}$} for more details about the orbital-resolved
  {$\mathrm{MAE}$}, energy band structures, {$\mathrm{Berry}$} curvature
  distributions, anomalous hall conductivity, and topological edge states of
  single-layer {${\mathrm{H-FeX}}_{2} {\mathrm{(X=Cl, Br, I)}}$}},}\
  }\BibitemShut {NoStop}%
\bibitem [{\citenamefont {Whangbo}\ \emph {et~al.}(2019)\citenamefont
  {Whangbo}, \citenamefont {Xiang}, \citenamefont {Koo}, \citenamefont
  {Gordon},\ and\ \citenamefont {Whitten}}]{doi:10.1021/acs.inorgchem.9b00687}%
  \BibitemOpen
  \bibfield  {author} {\bibinfo {author} {\bibfnamefont {M.-H.}\ \bibnamefont
  {Whangbo}}, \bibinfo {author} {\bibfnamefont {H.}~\bibnamefont {Xiang}},
  \bibinfo {author} {\bibfnamefont {H.-J.}\ \bibnamefont {Koo}}, \bibinfo
  {author} {\bibfnamefont {E.~E.}\ \bibnamefont {Gordon}}, \ and\ \bibinfo
  {author} {\bibfnamefont {J.~L.}\ \bibnamefont {Whitten}},\ }\href {\doibase
  10.1021/acs.inorgchem.9b00687} {\bibfield  {journal} {\bibinfo  {journal}
  {Inorganic Chemistry}\ }\textbf {\bibinfo {volume} {58}},\ \bibinfo {pages}
  {11854} (\bibinfo {year} {2019})}\BibitemShut {NoStop}%
\bibitem [{\citenamefont {Schaibley}\ \emph {et~al.}(2016)\citenamefont
  {Schaibley}, \citenamefont {Yu}, \citenamefont {Clark}, \citenamefont
  {Rivera}, \citenamefont {Ross}, \citenamefont {Seyler}, \citenamefont {Yao},\
  and\ \citenamefont {Xu}}]{2016Valleytronics}%
  \BibitemOpen
  \bibfield  {author} {\bibinfo {author} {\bibfnamefont {J.~R.}\ \bibnamefont
  {Schaibley}}, \bibinfo {author} {\bibfnamefont {H.}~\bibnamefont {Yu}},
  \bibinfo {author} {\bibfnamefont {G.}~\bibnamefont {Clark}}, \bibinfo
  {author} {\bibfnamefont {P.}~\bibnamefont {Rivera}}, \bibinfo {author}
  {\bibfnamefont {J.~S.}\ \bibnamefont {Ross}}, \bibinfo {author}
  {\bibfnamefont {K.~L.}\ \bibnamefont {Seyler}}, \bibinfo {author}
  {\bibfnamefont {W.}~\bibnamefont {Yao}}, \ and\ \bibinfo {author}
  {\bibfnamefont {X.}~\bibnamefont {Xu}},\ }\href
  {https://www.nature.com/articles/natrevmats201655} {\bibfield  {journal}
  {\bibinfo  {journal} {Nature Reviews Materials}\ }\textbf {\bibinfo {volume}
  {1}},\ \bibinfo {pages} {16055} (\bibinfo {year} {2016})}\BibitemShut
  {NoStop}%
\bibitem [{\citenamefont {Zhao}\ \emph {et~al.}(2019)\citenamefont {Zhao},
  \citenamefont {Ma}, \citenamefont {Lei}, \citenamefont {Wang},\ and\
  \citenamefont {Dai}}]{2019Single}%
  \BibitemOpen
  \bibfield  {author} {\bibinfo {author} {\bibfnamefont {P.}~\bibnamefont
  {Zhao}}, \bibinfo {author} {\bibfnamefont {Y.}~\bibnamefont {Ma}}, \bibinfo
  {author} {\bibfnamefont {C.}~\bibnamefont {Lei}}, \bibinfo {author}
  {\bibfnamefont {H.}~\bibnamefont {Wang}}, \ and\ \bibinfo {author}
  {\bibfnamefont {Y.}~\bibnamefont {Dai}},\ }\href
  {https://aip.scitation.org/doi/citedby/10.1063/1.5129311} {\bibfield
  {journal} {\bibinfo  {journal} {Applied Physics Letters}\ }\textbf {\bibinfo
  {volume} {115}},\ \bibinfo {pages} {261605} (\bibinfo {year}
  {2019})}\BibitemShut {NoStop}%
\bibitem [{\citenamefont {Li}\ \emph {et~al.}(2013)\citenamefont {Li},
  \citenamefont {Cao}, \citenamefont {Niu}, \citenamefont {Shi},\ and\
  \citenamefont {Feng}}]{doi:10.1073/pnas.1219420110}%
  \BibitemOpen
  \bibfield  {author} {\bibinfo {author} {\bibfnamefont {X.}~\bibnamefont
  {Li}}, \bibinfo {author} {\bibfnamefont {T.}~\bibnamefont {Cao}}, \bibinfo
  {author} {\bibfnamefont {Q.}~\bibnamefont {Niu}}, \bibinfo {author}
  {\bibfnamefont {J.}~\bibnamefont {Shi}}, \ and\ \bibinfo {author}
  {\bibfnamefont {J.}~\bibnamefont {Feng}},\ }\href {\doibase
  10.1073/pnas.1219420110} {\bibfield  {journal} {\bibinfo  {journal}
  {Proceedings of the National Academy of Sciences}\ }\textbf {\bibinfo
  {volume} {110}},\ \bibinfo {pages} {3738} (\bibinfo {year}
  {2013})}\BibitemShut {NoStop}%
\bibitem [{\citenamefont {Armitage}\ \emph {et~al.}(2018)\citenamefont
  {Armitage}, \citenamefont {Mele},\ and\ \citenamefont
  {Vishwanath}}]{RevModPhys.90.015001}%
  \BibitemOpen
  \bibfield  {author} {\bibinfo {author} {\bibfnamefont {N.~P.}\ \bibnamefont
  {Armitage}}, \bibinfo {author} {\bibfnamefont {E.~J.}\ \bibnamefont {Mele}},
  \ and\ \bibinfo {author} {\bibfnamefont {A.}~\bibnamefont {Vishwanath}},\
  }\href {\doibase 10.1103/RevModPhys.90.015001} {\bibfield  {journal}
  {\bibinfo  {journal} {Rev. Mod. Phys.}\ }\textbf {\bibinfo {volume} {90}},\
  \bibinfo {pages} {015001} (\bibinfo {year} {2018})}\BibitemShut {NoStop}%
\bibitem [{\citenamefont {Thouless}\ \emph {et~al.}(1982)\citenamefont
  {Thouless}, \citenamefont {Kohmoto}, \citenamefont {Nightingale},\ and\
  \citenamefont {Den~Nijs}}]{PhysRevLett.49.405}%
  \BibitemOpen
  \bibfield  {author} {\bibinfo {author} {\bibfnamefont {D.~J.}\ \bibnamefont
  {Thouless}}, \bibinfo {author} {\bibfnamefont {M.}~\bibnamefont {Kohmoto}},
  \bibinfo {author} {\bibfnamefont {M.~P.}\ \bibnamefont {Nightingale}}, \ and\
  \bibinfo {author} {\bibfnamefont {M.}~\bibnamefont {den~Nijs}},\ }\href
  {\doibase 10.1103/PhysRevLett.49.405} {\bibfield  {journal} {\bibinfo
  {journal} {Phys. Rev. Lett.}\ }\textbf {\bibinfo {volume} {49}},\ \bibinfo
  {pages} {405} (\bibinfo {year} {1982})}\BibitemShut {NoStop}%
\bibitem [{\citenamefont {Liu}\ \emph {et~al.}(2016)\citenamefont {Liu},
  \citenamefont {Zhang},\ and\ \citenamefont
  {Qi}}]{doi:10.1146/annurev-conmatphys-031115-011417}%
  \BibitemOpen
  \bibfield  {author} {\bibinfo {author} {\bibfnamefont {C.-X.}\ \bibnamefont
  {Liu}}, \bibinfo {author} {\bibfnamefont {S.-C.}\ \bibnamefont {Zhang}}, \
  and\ \bibinfo {author} {\bibfnamefont {X.-L.}\ \bibnamefont {Qi}},\ }\href
  {\doibase 10.1146/annurev-conmatphys-031115-011417} {\bibfield  {journal}
  {\bibinfo  {journal} {Annual Review of Condensed Matter Physics}\ }\textbf
  {\bibinfo {volume} {7}},\ \bibinfo {pages} {301} (\bibinfo {year}
  {2016})}\BibitemShut {NoStop}%
\bibitem [{\citenamefont {He}\ \emph {et~al.}(2018)\citenamefont {He},
  \citenamefont {Wang},\ and\ \citenamefont
  {Xue}}]{doi:10.1146/annurev-conmatphys-033117-054144}%
  \BibitemOpen
  \bibfield  {author} {\bibinfo {author} {\bibfnamefont {K.}~\bibnamefont
  {He}}, \bibinfo {author} {\bibfnamefont {Y.}~\bibnamefont {Wang}}, \ and\
  \bibinfo {author} {\bibfnamefont {Q.-K.}\ \bibnamefont {Xue}},\ }\href
  {\doibase 10.1146/annurev-conmatphys-033117-054144} {\bibfield  {journal}
  {\bibinfo  {journal} {Annual Review of Condensed Matter Physics}\ }\textbf
  {\bibinfo {volume} {9}},\ \bibinfo {pages} {329} (\bibinfo {year}
  {2018})}\BibitemShut {NoStop}%
\bibitem [{\citenamefont {Jungwirth}\ \emph {et~al.}(2002)\citenamefont
  {Jungwirth}, \citenamefont {Niu},\ and\ \citenamefont
  {MacDonald}}]{PhysRevLett.88.207208}%
  \BibitemOpen
  \bibfield  {author} {\bibinfo {author} {\bibfnamefont {T.}~\bibnamefont
  {Jungwirth}}, \bibinfo {author} {\bibfnamefont {Q.}~\bibnamefont {Niu}}, \
  and\ \bibinfo {author} {\bibfnamefont {A.~H.}\ \bibnamefont {MacDonald}},\
  }\href {\doibase 10.1103/PhysRevLett.88.207208} {\bibfield  {journal}
  {\bibinfo  {journal} {Phys. Rev. Lett.}\ }\textbf {\bibinfo {volume} {88}},\
  \bibinfo {pages} {207208} (\bibinfo {year} {2002})}\BibitemShut {NoStop}%
\bibitem [{\citenamefont {Yao}\ \emph {et~al.}(2004)\citenamefont {Yao},
  \citenamefont {Kleinman}, \citenamefont {MacDonald}, \citenamefont {Sinova},
  \citenamefont {Jungwirth}, \citenamefont {Wang}, \citenamefont {Wang},\ and\
  \citenamefont {Niu}}]{PhysRevLett.92.037204}%
  \BibitemOpen
  \bibfield  {author} {\bibinfo {author} {\bibfnamefont {Y.}~\bibnamefont
  {Yao}}, \bibinfo {author} {\bibfnamefont {L.}~\bibnamefont {Kleinman}},
  \bibinfo {author} {\bibfnamefont {A.~H.}\ \bibnamefont {MacDonald}}, \bibinfo
  {author} {\bibfnamefont {J.}~\bibnamefont {Sinova}}, \bibinfo {author}
  {\bibfnamefont {T.}~\bibnamefont {Jungwirth}}, \bibinfo {author}
  {\bibfnamefont {D.-s.}\ \bibnamefont {Wang}}, \bibinfo {author}
  {\bibfnamefont {E.}~\bibnamefont {Wang}}, \ and\ \bibinfo {author}
  {\bibfnamefont {Q.}~\bibnamefont {Niu}},\ }\href {\doibase
  10.1103/PhysRevLett.92.037204} {\bibfield  {journal} {\bibinfo  {journal}
  {Phys. Rev. Lett.}\ }\textbf {\bibinfo {volume} {92}},\ \bibinfo {pages}
  {037204} (\bibinfo {year} {2004})}\BibitemShut {NoStop}%
\bibitem [{\citenamefont {Whangbo}\ \emph {et~al.}(2015)\citenamefont
  {Whangbo}, \citenamefont {Gordon}, \citenamefont {Xiang}, \citenamefont
  {Koo},\ and\ \citenamefont {Lee}}]{doi:10.1021/acs.accounts.5b00408}%
  \BibitemOpen
  \bibfield  {author} {\bibinfo {author} {\bibfnamefont {M.-H.}\ \bibnamefont
  {Whangbo}}, \bibinfo {author} {\bibfnamefont {E.~E.}\ \bibnamefont {Gordon}},
  \bibinfo {author} {\bibfnamefont {H.}~\bibnamefont {Xiang}}, \bibinfo
  {author} {\bibfnamefont {H.-J.}\ \bibnamefont {Koo}}, \ and\ \bibinfo
  {author} {\bibfnamefont {C.}~\bibnamefont {Lee}},\ }\href {\doibase
  10.1021/acs.accounts.5b00408} {\bibfield  {journal} {\bibinfo  {journal}
  {Accounts of Chemical Research}\ }\textbf {\bibinfo {volume} {48}},\ \bibinfo
  {pages} {3080} (\bibinfo {year} {2015})}\BibitemShut {NoStop}%
\end{thebibliography}
\end{document}